\documentclass[preprint, 12pt]{aastex6}
\usepackage{amsmath,amssymb,bm, graphicx,xspace,multirow,threeparttable}
\usepackage{epsfig}
\usepackage{color}
\RequirePackage{lineno}
\usepackage{txfonts}
\usepackage{floatrow}
\def\asca       {{\em ASCA}\/}
\def\suzaku       {{\em Suzaku}\/}
\def\cha        {{\em Chandra}\/}

\def\xmm        {{\em XMM-Newton}\/}

\def\rosat      {{\em ROSAT}\/}

\makeatletter
\DeclareRobustCommand{\HI}{%
  \mbox{H\check@mathfonts\fontsize\sf@size\z@\selectfont\ I}%
}

\makeatother

\begin{document}

\title{Revealing a peculiar supernova remnant G106.3+2.7 as a petaelectronvolt proton accelerator with X-ray observations}

\author{Chong Ge$^{1}$, Ruo-Yu Liu$^{2,3,*}$, Shu Niu$^{4,5}$,
Yang Chen$^{2,3}$, Xiang-Yu Wang$^{2,3}$}
\affil{$^{1}$Department of Physics and Astronomy, University of Alabama in Huntsville, Huntsville, AL 35899, USA\\
$^{2}$School of Astronomy and Space Science, Nanjing University, Nanjing 210093, China;*Email:\textcolor{blue}{ryliu@nju.edu.cn}\\
$^{3}$Key laboratory of Modern Astronomy and Astrophysics (Nanjing University), Ministry of Education, Nanjing 210023, China\\
$^{4}$University of Chinese Academy of Sciences (CAS), Beijing 100049, China\\
$^{5}$Key Laboratory for Research in Galaxies and Cosmology, Shanghai Astronomical Observatory, CAS, 80 Nandan Road, Shanghai 200030, People’s Republic of China}

\begin{abstract}
Supernova remnants (SNRs) have long been considered as one of the most promising sources of Galactic cosmic rays. 
In the SNR paradigm, petaelectronvolt (PeV) proton acceleration may only be feasible at the early evolution stage, lasting a few hundred years, when the SNR shock speed is high. While evidence supporting the acceleration of PeV protons in young SNRs has yet to be discovered, X-ray synchrotron emission is an important indicator of fast shock. We here report the first discovery of X-ray synchrotron emission from the possibly middle-aged SNR~G106.3+2.7, implying that this SNR is still an energetic particle accelerator despite its age. This discovery, along with the ambient environmental information, multiwavelength observation, and theoretical arguments, supports SNR~G106.3+2.7 as a likely powerful PeV proton accelerator.
\end{abstract}

\maketitle

Cosmic rays (CRs) are high-energy charged particles moving through space at almost the speed of light. They serve as free experimental samples (with a composition of 90\% protons, 9\% helium nuclei, and 1\% heavier elements) to study particle physics and high-energy astrophysics. However, it is impossible to localise the origin of CRs from direction arrival directions because they are charged particles whose paths have been deflected by magnetic fields intervening between their sources and Earth. Where and how CRs are accelerated remain open questions.
The local CR proton spectrum roughly maintains a power-law shape up to the so-called `knee' around $1\,$PeV ($=10^{15}$eV), indicating the existence of powerful proton petaelectronvolt accelerators (PeVatrons) residing in our Galaxy.
However, despite decades of efforts, no Galactic sources have been firmly identified as proton PeVatrons except for the indication of a candidate at the Galactic Center \citep{HESS16_GC}. In the supernova remnant (SNR) paradigm of the origin of Galactic CRs, the maximum achievable particle energy is sensitive to the speed of SNR shock. Thus, only SNRs younger than a few hundred years old are considered capable of accelerating PeV protons\citep{Schure13,Bell15}. However, observational evidence has yet to be discovered.

\begin{figure}[ht!]
\centering
\includegraphics[width=0.8\textwidth,keepaspectratio=true,clip=true]{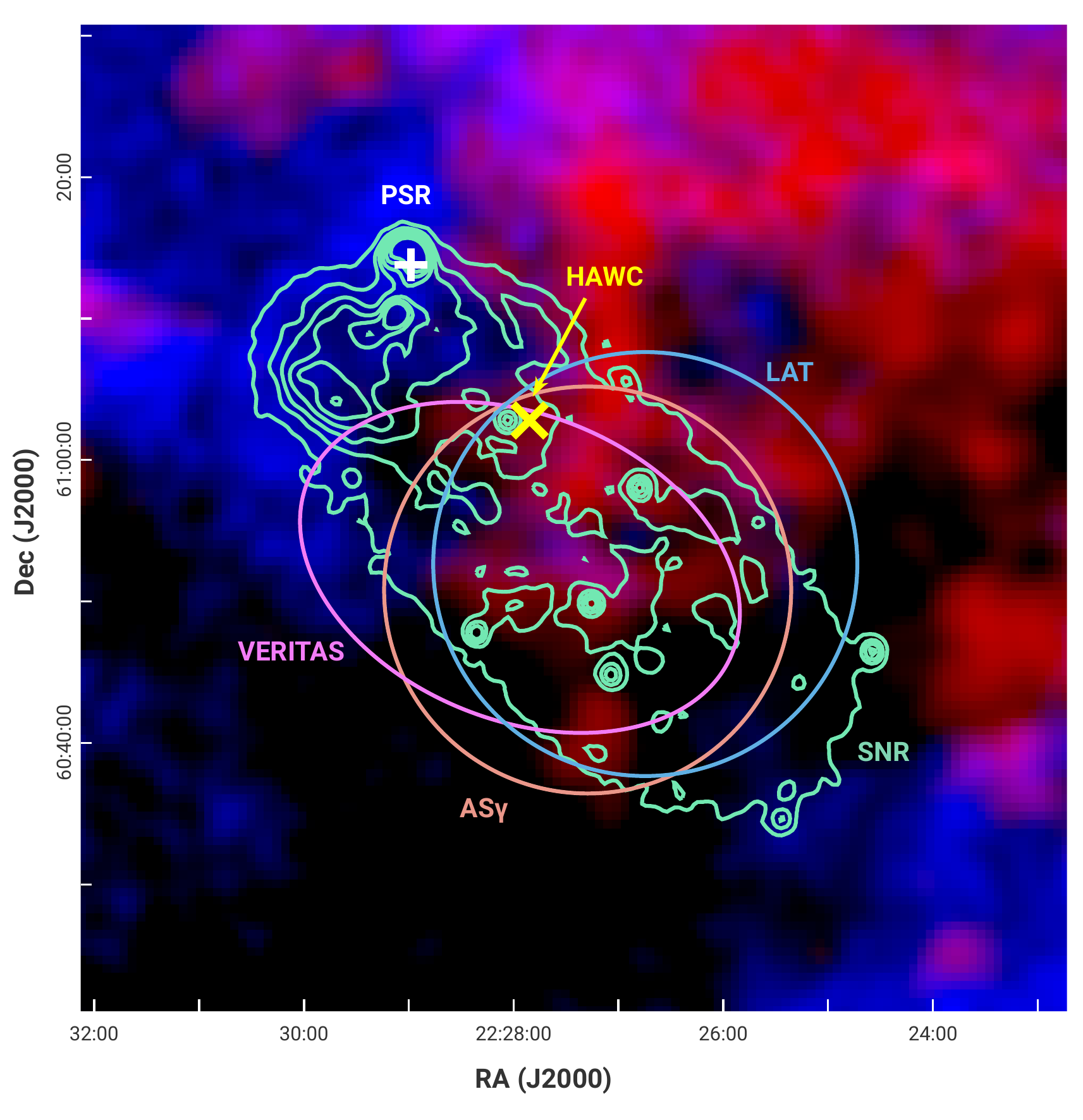}
	\caption{{\bf Gas Around SNR~G106.3+2.7} Blue color represents the distribution of \HI\ gas, and red color shows the CO \citep{Kothes01} emission of a dense molecular cloud. The green contours are the 1.4 GHz radio continuum of the SNR. The white plus marks the PSR~J2229+6114. The magenta ellipse, cyan circle, and orange circle represent the gamma-ray emission detected by VERITAS, Fermi-LAT, and Tibet AS$\gamma$, respectively. The yellow cross shows the best-fit position of the HAWC source.  RA and Dec. are in J2000.
}
\label{fig:gas}
\end{figure}

SNR~G106.3+2.7 was originally identified in a radio survey of the Galactic plane \citep{Joncas90}. As shown in Figure~\ref{fig:gas}, it appears as a cometary structure with a compact `head' of high surface brightness located in the northeastern part of the system and an elongated `tail' of low surface brightness extending toward the southwest \citep{Pineault00}. The northern part of the `head' region is associated with the `Boomerang' pulsar wind nebula (PWN), which is powered by the energetic pulsar PSR~J2229+6114 with a high spin-down luminosity of $2.2\times 10^{37}\rm erg\,s^{-1}$. The pulsar's characteristic age, i.e., 10\,kyr, is usually considered as a representation of the system's age \citep{Halpern01b,Kothes06}, although the true age of the pulsar could be younger (Figure~\ref{fig:age}). The PWN is enveloped by a small \HI\ shell, implying that the PWN is either pushing the \HI\ gas outward or ionizing the atomic hydrogen gas in its vicinity \citep{Kothes01}. A dense, shell-like \HI\ structure is also present in the southeastern boundary of the head region, this is likely caused by the encounter of the SNR shock with the dense ambient medium. This scenario is supported by the shell-like enhanced radio emission in the southeastern part of the head region. Based on the central velocity of the \HI\ emission, the distance of the system is considered as $d=800$\,pc \citep{Kothes01}, whereas a larger distance is suggested by \citet{Halpern01a}.
In contrast, the tail region appears to be expanding in a low-density hydrogen bubble that may have been excavated by the stellar winds of the previous generation of massive stars \citep{Kothes01}.

In the last decade, this region has attracted a lot of attention from the gamma-ray community\citep{Milagro07,Milagro09,Veritas09,Xin19,HAWC20,2021NatAs.tmp...41T}. Milagro detected TeV emission up to 35\,TeV with a large spatial uncertainty covering the entire system. VERITAS \citep{Veritas09} and Fermi-LAT \citep{Xin19} detected gamma-ray emission in the range of $0.9-16\,$TeV and $3-500\,$GeV, respectively, from the tail region. More recently, 
the HAWC \citep{HAWC20} and Tibet AS$\gamma$ \citep{2021NatAs.tmp...41T} experiments extended the spectrum up to 100\,TeV with consistent spatial position of the source from the VERITAS and Fermi-LAT regions as shown in Figure~\ref{fig:gas}. The AS$\gamma$ data indicates that the measured emission centroid deviates from the pulsar's location, with a confidence level of $3.1\sigma$\citep{2021NatAs.tmp...41T}, though one cannot rule out the possibility that a small fraction of the emission originates from the head region. Such measurements indicate that very energetic particles are present in the tail region. In principle, both multi-hundred TeV electrons and PeV protons can produce hundred-TeV gamma-ray photons \citep{Veritas09,Xin19,HAWC20,LiuSM20}, but the predicted X-ray emission is different in the leptonic and hadronic scenarios.

The Boomerang Nebula is a bright point source surrounded by diffuse emission in the X-ray band with a hard nonthermal spectrum of the photon index $1.5$ \citep{Halpern01a,Halpern01b}. It shows a centrally peaked morphology in the X-ray band, with 29\% of the emission coming from PSR~J2229+6114. The emission is largely confined within a radius of 100$''$, spatially coinciding with the shell structure in the radio band, as revealed by an early research \citep{Halpern01a,Halpern01b} with $\sim 10\,$ks \cha\ exposure, as well as by the observations of \asca\ and \rosat. Currently, the X-ray data from the region of the SNR-PWN complex are much more abundant (summarized in Table.~\ref{t:obs}), with exposure more than 100\,ks by \cha, \xmm, and \suzaku, respectively.

\begin{figure}
\centering
\includegraphics[width=0.8\textwidth,keepaspectratio=true,clip=true]{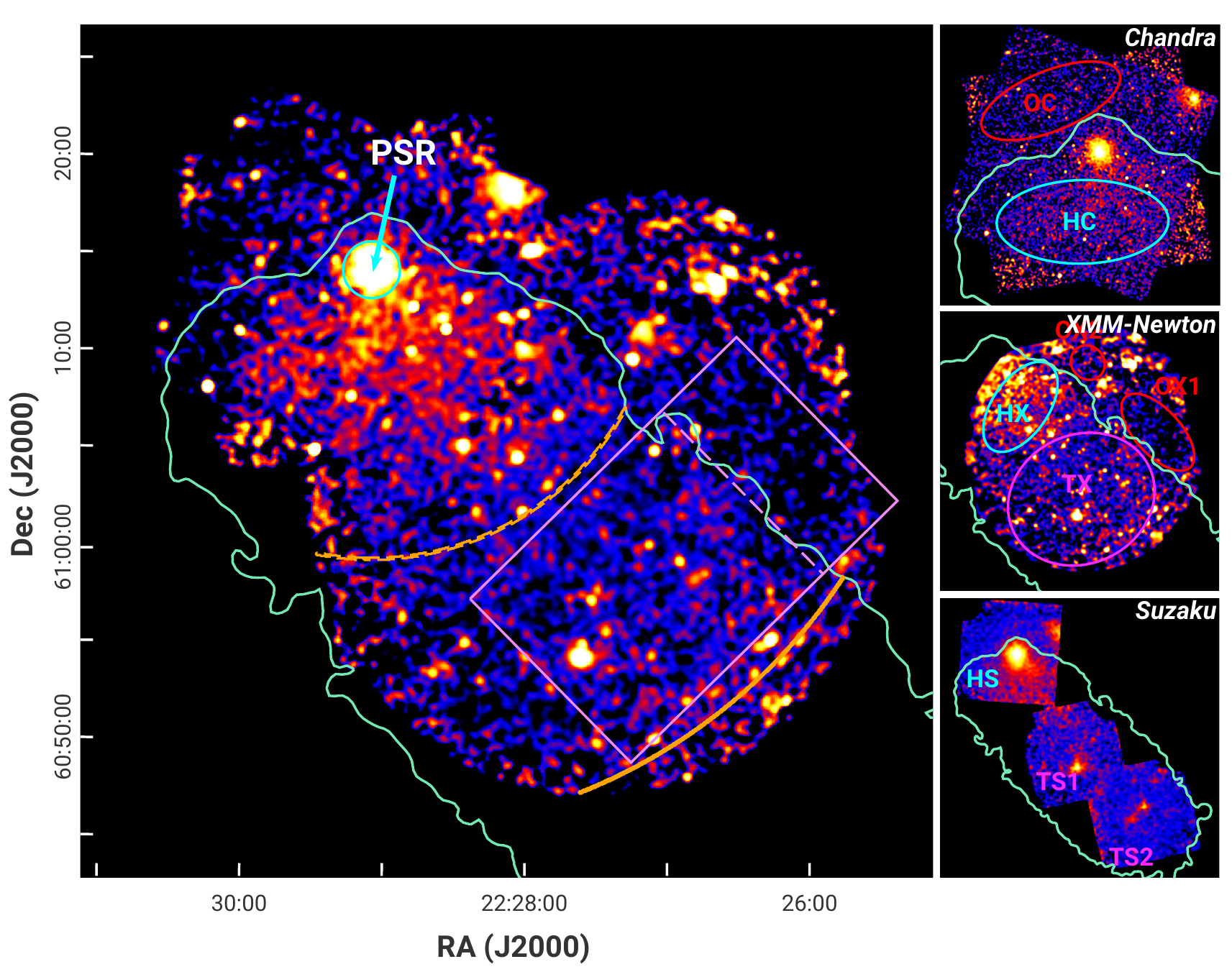}
	\caption{{\bf X-ray Images of SNR~G106.3+2.7}
Combined \cha\ and \xmm\ image in $1-7$\,keV with instrumental background subtracted and exposure corrected. 
The green curve outlines the SNR observed by the 1.4\,GHz radio continuum \citep{Pineault00}. The cyan circle has a radius of 100$''$ centered at the position of PSR~J2229+6114, characterizing the boundary of the Boomerang Nebula. The dashed orange arc marks the approximate boundary between the head and the tail of the SNR in the X-ray band, based on the SBP extracted from a series of annuli sectors centered at PSR~J2229+6114 out to the position marked with the solid orange arc (see Figure~\ref{fig:sbp}A). The magenta box is the intensity extraction region that shows the brightness change at the boundary of the SNR (see Figure~\ref{fig:sbp}B). 
The three small panels show respectively \cha, \xmm, and \suzaku\ images in 1-7 keV with marked regions for spectral analysis in Table~\ref{t:spec}. H is short for the head, T is for the tail, and O is for outside the SNR; C is for \cha, X is for \xmm, and S is for \suzaku. The same green contour outlines the SNR.
}
\label{fig:xray}
\end{figure}

\begin{figure}
\centering
\includegraphics[width=0.8\textwidth]{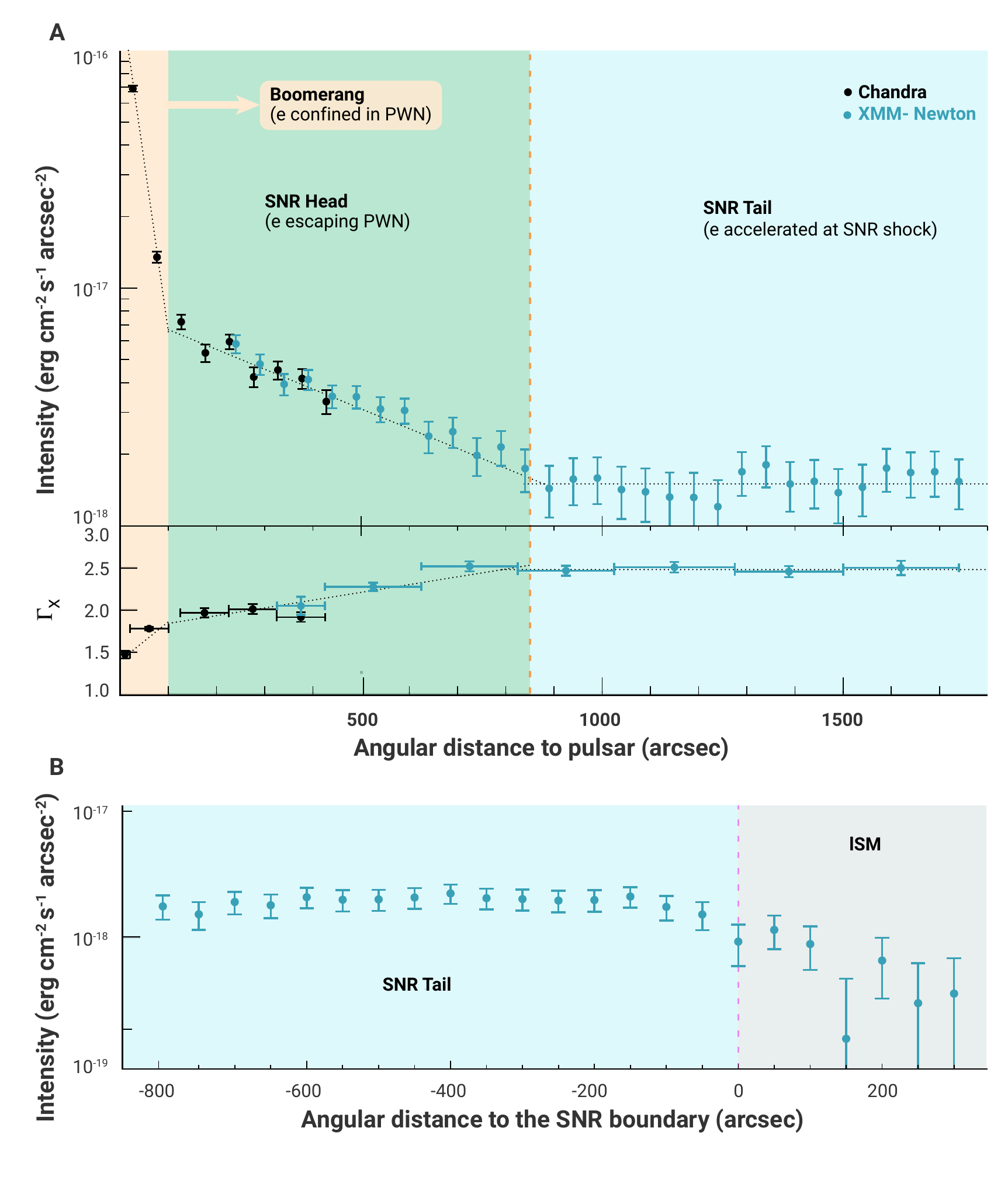}
	\caption{{\bf X-ray Surface Brightness Profiles} (A) Radial profiles of X-ray surface brightness and the spectral index extracted from a series of annuli centered at PSR~J2229+6114. The central bright point source (i.e. the pulsar's emission) has been subtracted. The three segments of dotted lines represent a PWN component, an escaping electron component in the head region, and an SNR component in the tail region, respectively. The dashed orange line marks the approximate boundary between the head and the tail, corresponding to the position marked with the dashed orange arc in Figure~\ref{fig:xray}. (B) A one-dimensional X-ray SBP inside the magenta rectangle is shown in Figure~\ref{fig:xray} after averaging over the short edge of the rectangle. 
	The dashed magenta line marks the boundary that separates the SNR tail and the interstellar medium (ISM), corresponding to the position marked also with the dashed magenta line in Figure~\ref{fig:xray}.}
\label{fig:sbp}
\end{figure}

We analyze the X-ray data from these instruments in the relevant regions as shown in Figure~\ref{fig:xray} and find the X-ray emission in all the regions of the SNR dominated by a nonthermal component, with a spectrum being consistent with a featureless power-law function (see Figure~\ref{fig:spec} and Table.~\ref{t:spec}). No convincing evidence of line features was identified in the spectral data. The analysis of the Boomerang region in this study (with ten times more \cha\ exposure) confirms the results of the previous study, and detects fainter nonthermal X-ray emission permeating the entire head region outside the PWN. Similar to the X-ray surface brightness profile (SBP) inside the Boomerang Nebula that is within 100$''$ from the pulsar, the average radial profile of the extended component in the SNR head
decreases with the distance from the pulsar, although in a shallower manner, and is accompanied by a gradual softening of the spectrum. Such features are also observed in other PWNe in the
X-ray band \citep{Slane04,Chen06,Tsujimoto11,vanEtten11}, suggesting that the extended X-ray emission in the head region is related to the Boomerang Nebula and probably produced by the electrons escaping the PWN. The decline in the X-ray surface
brightness and softening in the spectral index are maintained roughly until the boundary between the SNR head and the tail, as shown by the orange dashed arc in Figure~\ref{fig:xray}. They then become constant in the tail region, as measured by \xmm.
The transition indicates that the X-ray-emitting electrons have a different origin in the tail region compared with those in the head region. Since the X-ray-emitting electrons cool rapidly, these electrons are accelerated \textit{in situ}. The most plausible accelerator in the tail region is the SNR shock. This scenario is corroborated by the fact that a decline in brightness begins immediately outside the northwest boundary of the SNR tail, as shown in Figure~\ref{fig:sbp}B (see also the intensity contrast of the TX, TS1, and TS2 regions with the OX1 and OX2 regions in Table.~\ref{t:spec}), demonstrating that fresh X-ray-emitting electrons are produced inside the SNR.

\begin{table}[htbp]
\centering
  \begin{tabular}{@{}lccc@{}}
\hline\hline
Reg$^a$ & PL index & {Intensity}$^b$ (erg cm$^{-2}$ s$^{-1}$ arcmin$^{-2}$) & $\chi^2$/DOF \\ 
\hline
PWN & $1.7\pm0.1$ & $(1.1\pm0.1)\times10^{-13}$ & 307.6/303\\
\hline
HC & $1.9\pm0.1$ & $(1.3\pm0.1)\times10^{-14}$ & 734.9/606\\
HX & $2.2\pm0.1$ & $(1.3\pm0.1)\times10^{-14}$ & 230.7/192\\
HS & $2.0\pm0.1$ & $(1.2\pm0.1)\times10^{-14}$ & 207.6/240\\
\hline
TX & $2.4\pm0.1$ & $(5.4\pm0.5)\times10^{-15}$ & 1122.0/892\\
TS1 & $2.0\pm0.1$ & $(7.2\pm0.4)\times10^{-15}$ & 98.8/113\\
TS2 & $2.0\pm0.1$ & $(5.7\pm0.3)\times10^{-15}$ & 181.5/166\\
\hline
OC & $2.6\pm0.5$ & $(1.3\pm0.9)\times10^{-15}$ & 399.6/364\\
OX1 & $4.5\pm0.7$ & $(1.0\pm0.8)\times10^{-15}$ & 224.6/185\\
OX2 & $5.8\pm1.2$ & $(1.9\pm1.8)\times10^{-15}$ & 156.9/134\\
\hline
{BKG} & - & $(5.6\pm1.0)\times10^{-15}$ & -\\
\hline
\end{tabular}
  \caption{\bf Spectral Properties in Different Regions of SNR~G106.3+2.7.} 
  \vspace{0.3cm}
 \begin{tablenotes}
  \item $^a$PWN represents the emission within 100$''$ of the pulsar but excluding the emission of the pulsar itself. H stands for the head, T for the tail, and O for outside SNR; C stands for \cha, X for \xmm, and S for \suzaku. The corresponding spectral extraction regions are marked in Figure~\ref{fig:xray}. BKG stands for sky background intensity.
  \item $^b$ Intensity in 1-7 keV.
  \end{tablenotes}
\label{t:spec}
\end{table}

X-ray synchrotron emission has been widely considered as an important indicator of efficient particle acceleration. Previous studies\citep{Koyama95,Koyama97,Slane97,Bamba00,Gotthelf01,Hwang02,Reynolds07,Reynolds08} have demonstrated that X-ray synchrotron emission only appears in young SNRs (a few thousand years old) in which the SNR shock speed $v_s$ exceeds a few thousand kilometers per second. From a theoretical point of view, the particle acceleration timescale for a particle of energy $E$ (predicted by the diffuse shock acceleration theory \citep{Bell78,Blandford87}) strongly depends on the shock speed. It can be estimated by \citep{Kirk01, Rieger07}
\begin{equation}\label{eq:tacc}
    t_{\rm acc}\approx \frac{3D}{v_s^2}=\eta^{-1} \left(\frac{r_g}{c}\right)\left(\frac{c}{v_s}\right)^2\simeq 4 \eta^{-1}\left(\frac{E}{1\rm TeV}\right)\left(\frac{B}{10\mu\rm G}\right)^{-1}\left(\frac{v_s}{3000\rm km~s^{-1}}\right)^{-2}\,\rm yr 
\end{equation}
where $D=\eta^{-1}r_gc/3$ is the spatial diffusion coefficient of the particle, $r_g$ is the particle's Larmor radius, $c$ is the speed of light, $B$ is the magnetic field near the shock front, and $\eta(\leq 1)$ represents the deviation of the diffusion coefficient from the `Bohm diffusion'. This study considers it as the acceleration efficiency. The acceleration of electrons is generally limited by the cooling of synchrotron radiation. Equating the acceleration timescale to the synchrotron cooling timescale, the maximum achievable electron energy in the cooling-limited scenario is obtained and the synchrotron spectral cutoff energy is subsequently derived \citep{Parizot06,Zirakashvili07}, i.e.,
\begin{equation}\label{eq:v_emax}
    \epsilon_{\rm syn, max}\approx 7 \eta(v_s/3000{\rm km~s^{-1}})^2\, \rm keV,
\end{equation}
which is only dependent on the shock speed and $\eta$. The continuation of the X-ray spectrum up to 7\,keV without a clear spectral cutoff suggests that the SNR shock in the tail region has a high speed that is $v_s\geq 3000 \eta^{-1/2} \rm km/s$. 

In general, the inferred high shock speed is unlikely for a middle-aged SNR. However, considering that the tail region is formed by the SNR shock breaking out into a low-density cavity, it is possible that the shock is not significantly decelerated and maintains a high speed. Specifically, the SNR in the tail region may not have entered the Sedov-Taylor phase. In addition, the projected length of the tail region is $L_{\rm t}\simeq 14(d/800\,\rm pc)\,$pc {assuming the supernova explosion center is close to the position of PSR~J2229+6114} \citep{Kothes01}. Therefore, the mean projected shock speed in the tail region can be inferred independently as $v_s\sim L_{\rm t}/T_{\rm age}\simeq 1500\,{\rm km~s^{-1}}(d/800\,{\rm pc})(T_{\rm age}/10\,\rm kyr)^{-1}$. This velocity is generally consistent with the estimate from the X-ray spectrum. Note that the true shock velocity can be several times higher since the velocity inferred in this way is only the component projected on the plane of the sky. Additionally, if a shorter age and a larger distance of the system are considered, a larger mean velocity may be obtained.

The high shock speed empowers the tail region to accelerate high-energy protons. Unlike electron acceleration, the proton acceleration is age-limited \citep{Bell13} because the energy loss rate of protons is much slower than that of equal-energy electrons. Substituting the formulae for the maximum synchrotron energy $\epsilon_{\rm syn,max}$ or Eq.~(\ref{eq:v_emax}) into the formula of the acceleration timescale, the maximum proton energy achievable in the SNR shock of a dynamical age of $T_{\rm age}$ can be given by
\begin{equation}
    E_{p,\rm max}\approx 3 \left(\frac{T_{\rm age}}{10\,\rm kyr}\right)\left(\frac{B}{10\mu \rm G}\right)\left(\frac{\epsilon_{\rm syn, max}}{7\,\rm keV}\right)\, \rm PeV.
\end{equation}
In principle, the tail region may serve as a proton PeVatron as long as the magnetic field is not too weak. Generally, the magnetic field in an SNR shock is not expected to be weaker than that in ISM of a typical total strength of $5-10\,\mu$G \citep{Beck15,Han17}. 

\begin{figure}[ht]
    \centering
    \includegraphics[width=0.8\textwidth]{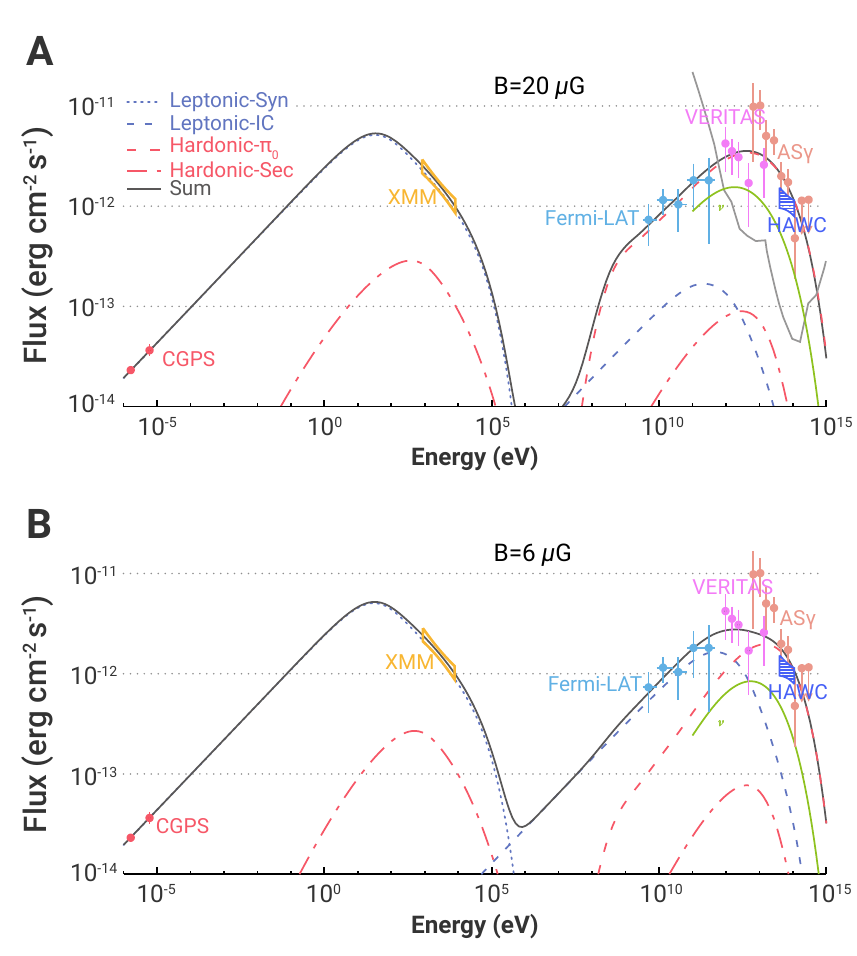}
    \caption{{\bf Modeling the SED of the Tail Region} Since the magnetic field is not clearly known, the magnetic fields of $B=20\mu$G (panel A) and $B=6\mu$G (panel B) are considered. In both panels, the dotted blue curves represent the synchrotron radiation, and the dashed blue curves represent the IC radiation of electrons. The dashed red curves show the pionic gamma-ray emission from $pp$ collisions of CR protons while the dot-dashed red curves show the emission of secondary electrons and positrons co-produced in the $pp$ collisions. The thick solid black curves show the sum of all these components. The green-yellow curves represent the expected $\nu_\mu+\bar{\nu}_\mu$ flux assuming a flavor ratio of 1:1:1 after oscillation. Model parameters for the case of $B=20\,\mu$G: total electron energy $W_e=3\times10^{46}$ergs, electron spectral break energy $E_{e,b}=5\,$TeV, the maximum electron energy $E_{e,\rm max}=200\,$TeV, total proton energy $W_p=10^{47}\,$ergs, proton spectral break energy $E_{p,b}=50\,$TeV, proton spectral index before the break $\alpha_p=1.7$, and the index change after the break $\Delta \alpha_p=0.5$. Model parameters for the case of $B=6\,\mu$G: $W_e=2.3\times10^{47}$ergs, $E_b=9\,$TeV, {$E_{e,\rm max}=400\,$TeV}, $W_p=4\times10^{46}\,$ergs, $E_{p,b}=300\,$TeV, $\alpha_p=1.6$, and $\Delta \alpha_p=0.7$. For both these cases, the electron spectral index before the break is $\alpha_e=2.3$, the index change after the break is $\Delta \alpha_e=1.4$, and the maximum proton energy is $E_{p,\rm max}=1\,$PeV. The gas density for the hadronuclear interaction is assumed to be $50\,\rm cm^{-3}$. The CGPS data follows \citet{Pineault00}. The Fermi-LAT data is analyzed by \citet{Xin19}. The VERITAS data is taken from \citet{Veritas09} and scaled up by a factor of 1.7 to account for the spillover effects \citep{HAWC20}. The HAWC data is taken from \citet{HAWC20}. The Tibet AS$\gamma$ data is taken from \citet{2021NatAs.tmp...41T}. The X-ray bowtie is obtained in this work by multiplying the intensity (flux per solid angle) of the TX region (Table.~\ref{t:spec}) with a solid angle of approximately $600\,\rm arcmin^2$ for the entire tail region. For reference, the one-year point-source sensitivity of LHAASO \citep{LHAASO19} is shown with the solid grey curve in panel (A).}
    \label{fig:sed}
\end{figure}

On the other hand, the spectral energy distribution (SED) of the tail region provides information on the magnetic field and even on the necessity of the acceleration of PeV protons in the tail region. In the modeling of the spectrum, $d=800\,$pc and $T_{\rm age}=10\,$kyr are employed as the fiducial parameters. For middle-aged SNRs, the emitting electron spectrum is usually given by a broken-power-law function with a high-energy cutoff \citep{Ohira17,ZhangX19}. We followed this description and found that when the radio and X-ray data in the tail region is well reproduced with the synchrotron radiation of electrons, the gamma-ray spectrum cannot be explained solely by electrons of the same population and the GeV data disfavors a magnetic field much weaker than $B=6\mu$G. As shown in Figure~\ref{fig:sed}, the GeV-TeV gamma-ray emission may or may not be reproduced by the inverse Compton (IC) radiation of electrons, depending on the assumed magnetic field. However, the spectrum beyond 30\,TeV cannot be interpreted by the IC radiation, because the Klein-Nishina effect suppresses the cross-section of IC scattering at high energies and softens the spectrum (Figure~\ref{fig:sed_B}).

As a result, a hadronic gamma-ray component that arises from the interactions between accelerated CR protons and gas must be introduced to explain the spectrum of the tail region at least above 10\,TeV, because there are no other reasonable particle accelerators in this region. In this scenario, the protons need to be accelerated to 1\,PeV or higher to explain the detection of 100\,TeV photons. The tail region is spatially coincident with a dense molecular cloud (Figure~\ref{fig:gas}) with an average atom density of $50\,\rm cm^{-3}$ \citep{Kothes01,HAWC20}, which can serve as the target for the hadronuclear interactions of CR protons escaping from the SNR. The gamma-ray spectrum can be interpreted well with a proton spectrum of a broken-power-law with a high-energy exponential cutoff. The spectral break may be caused by the CR confinement in the SNR because lower-energy CRs are more difficult to escape. Therefore, the spectrum becomes harder than the one accelerated at the shock. In contrast, CRs above a certain energy level can more easily escape the SNR, so the spectrum above the break energy is close to the theoretical expectation of shock acceleration with a power-law spectral index $\gtrsim 2$ \citep{Bell78,Blandford87}. The required total energy for escaping CRs that are interacting with the molecular cloud is less than $10^{47}\,$ergs. This is reasonable as it constitutes only a tiny fraction of an SNR's energy budget. The flux of co-produced (anti-)muon neutrinos in the hadronuclear interactions would result in a detection of 0.4 track-like events above 50\,TeV for the 10-year operation of the IceCube neutrino telescope. This is consistent with the non-detection of neutrino events from the direction of the source \citep{IC20_PWN}.  

The results of this study indicate that the tail region of SNR~G106.3+2.7 is likely a long-sought source of PeV Galactic CRs. Even though it is probably already in the middle-age stage, the nonthermal X-ray radiation from the tail region indicates that its shock velocity still remains at several thousand kilometers per second. This velocity is conducive to the sustained acceleration of protons to PeV energy. The inferred high shock velocity is supported by the expansion of the tail into a low-density cavity. This is consistent with the elongated morphology of the tail. In addition, the multiwavelength spectrum data of the tail region further indicates the necessity of the hadronic component produced by energetic protons with a spectrum extending up to 1\,PeV or beyond. This scenario is corroborated by the presence of a dense molecular cloud that spatially coincides with both the tail region and the centroid of the gamma-ray emissions. On the other hand, from a theoretical point of view, SNRs expanding in stellar wind cavities have been suggested as promising accelerators of PeV protons \citep{Voelk88,Zirakashvili18} because a quasi-perpendicular SNR shock may form in this environment, with an acceleration efficiency comparable to that in the Bohm limit (i.e., $\eta\sim 1$).  These various arguments consistently support that the tail region of SNR~G106.3+2.7 is likely a Galactic proton PeVatron. In addition to the tail region, the highly asymmetric morphology of the SNR also leaves a clue to study the co-evolution of the SNR-PWN complex system. In-depth theoretical studies on the dynamics, the particle acceleration and transport processes would facilitate a comprehensive understanding of the system.

SNR~G106.3+2.7 is peculiar. To date, it is potentially the first and the only middle-aged SNR detected with X-ray synchrotron emission. It is located in a special environment that favors particle acceleration. Its shock velocity is much higher when compared with other middle-aged SNRs,  so a high particle acceleration rate is expected. It has experienced a much longer lifetime when compared with young SNRs that have high shock velocity, offering sufficient time for the CRs to be gradually accelerated up to PeV energies. This discovery challenges the present SNR paradigm of CR origin. In the future, multiwavelength observations by high-performance instruments such as LHAASO, CTA, and eROSITA will largely promote the discovery and identification of more PeVatrons. Hopefully, these observations will unravel the century-old puzzle of the origin of Galactic CRs.

\vspace{20pt}

\noindent \textbf{\large Materials and Methods}\\

\noindent \textbf{X-ray Data Analysis}\\
\label{s:obs}
Parts of head and tail regions of SNR G106.3+2.7 are observed by \cha, \suzaku, and \xmm, respectively, which are listed in Table~\ref{t:obs}.
We process the \cha\ data using the \cha\ Interactive Analysis of Observation (CIAO; version 4.11) with calibration database (CALDB; version 4.8.2), and the \xmm\ data with the Extended Source Analysis Software (ESAS) that integrated into the \xmm\ Science Analysis System (SAS; version 17.0.0), following the procedures in \citet{2019MNRAS.484.1946G, 2020MNRAS.497.4704G}.
Since the \xmm\ observations were originally proposed for the Solar Wind Charge Exchange (SWCX) X-ray emission, we remove the X-ray emission below 1 keV that may be highly affected by the SWCX emission.
\suzaku\ data are reprocessed with the latest CALDB by the standard pipeline tool in {\sc heasoft} v6.26. For ease, we only analysis the \suzaku\ data in 3x3 mode and abandon the short exposure data in 5x5 mode.

In Fig.~\ref{fig:xray}, we combine the processed \cha\ and \xmm\ images as follow steps: 1) downgrade the resolution of \cha\ image to match the one of \xmm\ image; 2) rescale the \cha\ image to the \xmm\ image according to the ratio of these two images in their common region; 3) add these two images
and take the mean value in their common region. 
We do not combine with \suzaku\ image, because the spatial resolution of \suzaku\ ($\sim 2^{\prime}$) is much lower than \cha\ ($\sim 1^{\prime\prime}$) and \xmm\ ($\sim 5^{\prime\prime}$).
We overlaid the green contour from the 1.4 GHz radio continuum measured by CGPS to outline the morphology of SNR~G106.3+2.7. 
We notice some distinct contrasts of surface brightness among different regions as shown in Table~\ref{t:spec}. We therefore conduct a more detailed analysis on the X-ray SBP, paying particular attention to the transition of the surface brightness profile in different regions. As shown in Fig.~\ref{fig:xray}, the orange dashed arc roughly divides the head and tail regions. The magenta dashed line marks the boundary of the SNR in the tail region, with respect to ISM.  
We then extract SBPs in a series of consecutive annulus sectors centred at the pulsar within the SNR, excluding the X-ray point sources.
{We choose an annulus width of 50$^{\prime\prime}$ to include enough counts in each annulus for statistics.}
We can see from the upper panel of Fig.~\ref{fig:sbp} the change of the trend of the SBP on the boundary between the Boomerang Nebula and the SNR head, as well as on the boundary between the SNR head and the SNR tail. We also extract the SBPs on both sides of the northwest boundary of the SNR tail as marked with the dashed magenta line in Fig.~\ref{fig:xray}, with dividing the region of the magenta box into many rectangular slices perpendicular to the direction shown with the magenta arrow. The result is shown in the lower panel of Fig.~\ref{fig:sbp}.

To obtain the spectral information in different parts of the SNR, we extract spectra from \cha, \xmm, and \suzaku\ in regions (with X-ray point sources masked) as shown in Fig.~\ref{fig:xray}. Due to the limited resolution of \suzaku, its spectra are extracted from individual observations (marked as HS, TS1, and TS2) with detected bright point sources masked.
We also extract a diffuse sky background spectrum from the {\em ROSAT} All-Sky Survey (RASS) in a 1-2 degree annulus surrounding the SNR. 
We jointly fit the spectral data of RASS with that from \cha\/, \xmm\/, and \suzaku\ respectively to link the sky background in different analyses, which is modelled by two thermal components for the foreground Galactic soft X-ray emission and one power-law component for the unresolved cosmic X-ray background {(the intensity of sky background is listed in Table~\ref{t:spec})}.
The best-fit spectra from these regions {inside the SNR} are consistent with a featureless power-law (PL) shown in Fig.~\ref{fig:spec} and summarized in Table~\ref{t:spec}, {and the best-fit hydrogen column density is $N_{\rm H}\sim0.95\times 10^{22}{\rm\ cm}^{-2}$}.
{On the other hand, the X-ray emission outside the SNR is insignificant and consistent to the sky background fluctuation, although it might also arise from the radiation of escaping electrons from the PWN or the SNR. We then also test our results by replacing the RASS sky background with the nearby \xmm\ local background region (OX1) and find that using of local background gives comparable spectral results for the regions inside the SNR (e.g. the X-ray intensity fitted with \xmm\ local background is a bit lower than the one fitted with the RASS background, but the difference is within $1~\sigma$ uncertainty).}

As an example, we try a thermal model ({\sc apec}; abundance fixed to Solar) to fit the \xmm\ spectra in the TX region, which returns a temperature of $kT=3.3\pm 0.2\,$keV, however, with a worse fitting statistics of $\chi^2$/DOF = 1186.1/892 (v.s. $\chi^2$/DOF=1122.0/892 for the PL model). Utilizing the Akaike Information Criterion (AIC) \citep{AIC74}, we define ${\rm AIC}=-2\ln L + 2k$ where the logarithm of the likelihood $\ln L=-\chi^2/2$, and $k$ is the number of model parameters which is 2 for both models. We therefore get ${\rm AIC_{\rm TH}} - {\rm AIC_{\rm PL}}\simeq 64$, indicating that the PL model gives a significantly better description of the X-ray spectrum than the thermal model. {Fitting an {\tt NEI} model to the TX region gives a $kT=3.3$ keV and an ionization parameter $n_{\rm e}t=3.1\times10^{13} {\rm\ s\ cm}^{-3}$, still with a worse fitting statistics of $\chi^2$/DOF = 1180.5/891 ($\Delta {\rm AIC}\simeq 60$) worse than the PL model). Moreover, the electron density is $n_{\rm e}=0.06 {\rm\ cm}^{-3}$ from the {\tt NEI} normalization, assuming a uniform density distribution with a LOS width of 6\,pc at a distance of 800\,pc. While the number of obtained $n_e$ is consistent with a low-density wind cavity, the resulting  ionization timescale would be $5\times 10^{14}\,$s, which is too large to be reasonable.
Including an additional thermal component (i.e. PL+{\sc APEC}) to fit the X-ray emission in the \xmm\ tail region (TX) along with a PL component does not improve the fitting statistics ($\chi^2$/DOF = 1121.9/890 for PL+{\sc APEC} v.s. $\chi^2$/DOF = 1122.0/892 for PL only), while the best-fit parameters for the thermal component are unphysical with an extremely high temperature of 58\,keV and an extremely low normalization, which is several magnitudes lower than the one of PL and hence can be ignored.} 
Besides, as shown in Fig.~\ref{fig:spec}, we do not find evidence for emission lines except for strong instrumental lines of of Al K$\alpha$ at 1.49\,keV and Si K$\alpha$ at 1.75\,keV in \xmm\ spectra. The spikes shown in the spectra are most likely fake line features from noise, because they are not repetitive {even for the spectra extracted from the same region} among different observations, and their peaks would have appeared broader if they were true emission lines given that the energy resolution of CCD spectra is low ($\sim$ 100 eV). {The non-detection of line features also disfavors the thermal or NEI model.}

{We also extract a radial spectral profile in a series of annulus sectors shown in Fig.~\ref{fig:sbp}. The annulus is wider than the one for SBP, because we need more counts to constrain spectral parameters. The width is 100$^{\prime\prime}$ (20$^{\prime\prime}$ and 80$^{\prime\prime}$ for the first and second annulii from the PWN) for \cha\ and 225$^{\prime\prime}$ (100$^{\prime\prime}$ for the first annulus overlapped with \cha) for \xmm.}
Note that in Table.~\ref{t:spec} the flux from \suzaku\ in the head region is a little lower than the one from \cha\ and \xmm, because the HS spectra also cover the low surface brightness region outside the SNR.
The flux of \suzaku\ in the tail region is higher than the one of \xmm, because some point sources are unresolved in \suzaku\ limited to its resolution. Moreover, most of the unresolved point sources are cosmic X-ray background sources with a typical PL index of $\sim\ 1.5$, which makes the best-fit PL index from \suzaku\ harder than the one from \xmm.
\\

\noindent \textbf{Modelling the Multiwavelength Spectrum in the Tail Region}\\
It has been shown that the time-integrated electron spectrum in a SNR can be well described by a broken-PL function with a high-energy cutoff  \citep{Ohira17, ZhangX19}
\begin{equation}
 \frac{dN_e}{dE_e}=N_{0,e} E_e^{-\alpha_e}\left[1+\left(\frac{E_e}{E_{e,b}} \right)^\sigma \right]^{-\Delta \alpha_e/\sigma}\exp\left[-\left(\frac{E_e}{E_{e,\rm max}} \right)^2 \right],
\end{equation}
where $\alpha$ is the spectral index before the break, $\Delta \alpha$ is spectral change after the break energy $E_{e,b}$, $\sigma$ characterizes the smoothness of the transition, and $E_{e,\rm max}$ represents the high-energy cutoff energy. {$E_{e,\rm max}$ is obtained by equating the cooling timescale of electrons (see Fig.~\ref{fig:timescale}) with the acceleration timescale given by Eq.~(\ref{eq:tacc}), which result in $E_{e,\rm max}\simeq 220\,(B/20\mu{\rm G})^{-1/2}(v_s/5000{\rm km~s^{-1}})$}. $N_{0,e}$ is a normalization factor which will be determined by the integration $\int E_e\frac{dN_e}{dE_e}dE_e=W_e$ over a lower limit of 5\,MeV up to an upper limit of 5\,PeV, with $W_e$ being obtained by fitting the data. We fix $\sigma=5$ following the fitting results for many other middle-aged SNRs  \citep{ZhangX19} and find $\alpha_e=2.3$ with $\Delta \alpha_e=1.4$ can reproduce the spectrum from radio to X-ray band. The magnetic field in the tail region $B$ is left as a free parameter. We consider the cosmic microwave background and the interstellar radiation field at the location of SNR~G106.3+2.7 modelled by \citet{Popescu17} as the target photons for the IC radiation. The interstellar radiation field can be approximately described by the superposition of three grey body radiation: a far infrared radiation field with temperature $T=30\,$K with an energy density of $0.4\rm eVcm^{-3}$, a near infrared radiation field with $T=500\,$K with an energy density of $0.2\rm eVcm^{-3}$ and an optical radiation field with $T=5000\,$K with an energy density of $0.4\rm eVcm^{-3}$. {Relativistic electrons also radiate in X-ray to gamma-ray band through the nonthermal bremsstrahlung process. The radiation efficiency is proportional to the gas density. Since the shock is propagating in the low-density cavity, this process is not important.} 

Note that the ratio between the synchrotron luminosity and the IC luminosity radiated simultaneously by the same population of electrons is proportional to the ratio between the magnetic energy density and the radiation energy density $U_B/U_{\rm ph}$. For the considered case here, $U_{\rm ph}$ is the sum of the aforementioned interstellar radiation field and CMB, which are approximately fixed. Consequently, on the premise of fitting the spectrum from radio to X-ray band via the synchrotron radiation of electrons, the flux of the corresponding IC radiation depends on the employed magnetic field. In Fig.~\ref{fig:sed} we can see that the IC flux can reproduce well the Fermi-LAT data for $B=6\,\mu$G. A weaker magnetic field may be in tension with the Fermi-LAT data and VERITAS data. In Fig.~\ref{fig:sed_B}, we show the IC radiation with $B=4\mu$G, which is probably too weak for the magnetic field in an SNR shock {(c.f., \citealt{LiuSM20} for a pure leptonic interpretation of the SED but with the X-ray spectrum of the Boomerang region obtained by \citealt{Halpern01b})}. As expected, the IC radiation with such a weak magnetic field overshoots all the Fermi-LAT data points and most VERITAS data points. Even in this extreme case, the IC component still cannot explain the measured flux above 30\,TeV. The softening of the IC flux at high energy is due to the Klein-Nishina effect instead of the cutoff in the electron spectrum, as demonstrated by the dashed red curve where an extremely {(and unphysically}) high cutoff energy with $E_c=5\,$PeV is employed. 

\begin{figure}[ht!]
\centering
\includegraphics[width=0.8\textwidth]{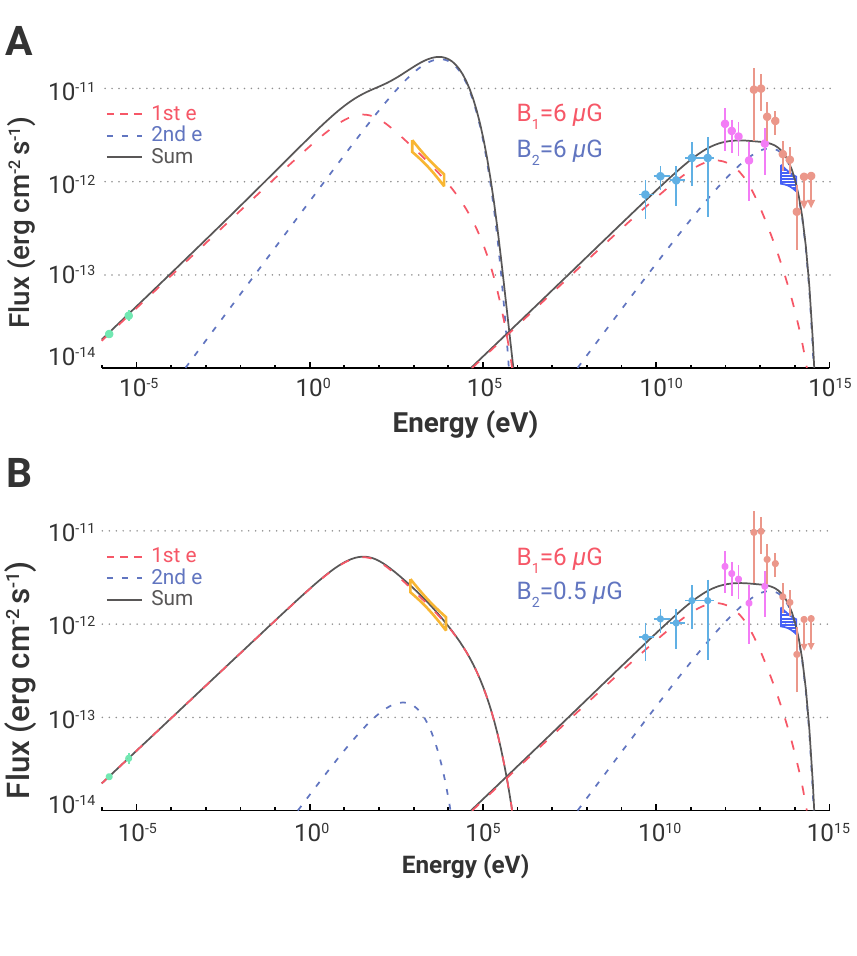}
\caption{{\bf Two-electron-component Scenario for the SED of the SNR Tail Region} The the red curves for the first electron component (`1st e') and the blue curves for the second electron component (`2nd e'). In both panels, the spectrum of the first electron component is identical to that in Figure~\ref{fig:sed}B while the spectrum of the second electron component is {$dN_{e,2}/dE_e\propto E_e^{-2}\exp \left[-(E_e/200\,{\rm TeV})^2\right]$ with a total energy $6\times 10^{45}\,$ergs}. The magnetic field employed for the synchrotron radiation of the second electron component is $B_2=6\,\mu$G in panel (A) while $B_2=0.5\,\mu$G is employed in panel (B). Data points are the same as those shown in Figure~\ref{fig:sed}.}
\label{fig:sed_2e}
\end{figure}

{We also check that whether the multiwavelength spectrum of the tail region can be interpreted by two electron components. In this case, the first electron component is supposed to explain the spectrum from radio to TeV gamma-ray, and is assumed to be the same as the leptonic component shown in the lower panel of Fig.~\ref{fig:sed} with a magnetic field $B_1=6\,\mu$G as required from the simultaneous fitting of the radio data and the GeV data. The second electron component is required to explain the spectrum beyond 30\,TeV and in the meanwhile not to overproduce the emission in other wavelength. To compensate for the KN effect beyond 30\,TeV, a hard spectrum for the second electron component is needed. Following this spirit, we find that employing $dN_{e,2}/dE_e \propto E_e^{-2}\exp\left[-\left(E_e/200{\rm TeV}\right)^2 \right]$ with $\int E_e dN_{e,2}/dE_e dE_e=6\times 10^{45}\,$ergs for the second component can successfully reproduce the gamma-ray spectrum. However, the synchrotron radiation of the second component dramatically overshoots the X-ray data if the same magnetic field for the first component, i.e., $B_2=B_1=6\,\mu$G is assumed, as is shown in the upper panel of Fig.~\ref{fig:sed_2e}. To make the expected X-ray flux consistent with the observation, the magnetic field has to be {tuned down to $B_2=0.5\,\mu$G} (see the lower panel of Fig.~\ref{fig:sed_2e}). It means that the second electron component is not from the same region of the first one which is in the SNR tail region, but what is the origin of the second electron component is probably a critical issue for this scenario. In addition, the required magnetic field is significantly weaker than that of the ISM. We therefore do not regard the two-electron-component scenario as a reasonable explanation to the multiwavelength data. }

Relativistic protons can produce gamma rays via the hadronuclear interaction, in which neutral pions are produced and further decay into gamma rays with each gamma-ray photon taking about 10\% of the parent proton's energy ($p+p\to p+p+\pi^0; \pi^0\to \gamma+\gamma$). In the modelling, the escaping proton spectrum is also assumed to be the same form of the electron spectrum, i.e.,
\begin{equation}
    \frac{dN_p}{dE_p}=N_{0,p} E_p^{-\alpha_p}\left[1+\left(\frac{E_p}{E_{p,b}}\right)^\sigma \right]^{-\Delta \alpha_p/\sigma}\exp\left(-\frac{E_p}{E_{p,\rm max}} \right).
\end{equation}
where $\sigma$ is the smooth of the spectral break and fixed at 2 in the calculation. The cutoff energy $E_{p,\rm max}$ is fixed at 1\,PeV in the calculation. Again, the normalization factor $N_{0,p}$ is obtained by $\int E_p\frac{dN_p}{dE_p}dE_p=W_p$. The pionic gamma-ray emission as well as the accompanying neutrino emission is calculated following the semi-analytical method developed by \citet{Kafexhiu14}. {The detection rate of neutrino events by the IceCube neutrino telescopes can be calculated by convolving the obtained neutrino flux (as shown in the dark yellow curves in Fig.~\ref{fig:sed}) and the effective area of IceCube (denoted by $A_{\rm eff}$) for the declination of the source ($30^\circ<\delta<90^\circ$) given by \citet{IC17}, i.e.
\begin{equation}
    N_{\nu_\mu+\bar{\nu}_\mu}=T_{\rm opr}\int\frac{dN_{\nu_{\mu}}}{dE_{\nu_\mu}}A_{\rm eff}(E_{\nu_{\mu}})dE_{\nu_\mu}
\end{equation}
where $T_{\rm opr}=10\,$yr is the operation time of IceCube.}
{In the meanwhile, secondary electrons/positrons are also produced with gamma rays in the hadronuclear interaction, with an emissivity about half of the gamma rays. We also calculate their synchrotron radiation in the magnetic field of ISM which is assumed to be $3\,\mu$G and the IC radiation in the interstellar radiation field, and find they have negligible contribution to the total flux.}

 {We note that although we assume both the spectra of electrons and protons to be a broken power law, the breaks might arise from different physics. The emitting electron are those that confined in the SNR shock so the spectrum before the break may represent the injection spectrum while the spectrum after the break could be softened by cooling (probably with the influence of a non-constant injection rate) or by the energy-dependent escape. By contrast, the emitting protons are those that escape the SNR and reach the molecular cloud. Therefore, the proton spectrum before the break are hardened by the confinement (which is opposite to the influence of escape) while the proton spectral index after the break may represents the injected one or perhaps is also modified by the diffusion in the molecular cloud or in the medium between the cloud and the SNR. Timescales for electron cooling and proton escape/confinement are shown in Fig.~\ref{fig:timescale}. We here simply choose the value of $\Delta\alpha_p$ so that the proton spectral index after the break (i.e., $\alpha_p+\Delta\alpha_p$) to be the same with the electron spectral index before the break which is about $\alpha_e=2.3$ from the fitting of the radio data. In fact, given the uncertainty of the gamma-ray flux around 100\,TeV, the proton spectral shape is not well constrained from the point of view of fitting. For the broken power-law function, $\alpha_p+\Delta\alpha_p$ in $2.0-2.4$ can give satisfactory fitting. One may also use a single PL function with a high-energy exponential cutoff for the proton spectrum to fit the gamma-ray data.}

\vspace{20pt}
\noindent \textbf{\large Supplemental Information}\\
Supplemental Material includes five figures, one table and some relevant discussions.

\vspace{20pt}
\noindent \textbf{\large Supplemental Material}\\

\setcounter{figure}{0}
\renewcommand{\thefigure}{S\arabic{figure}}

\setcounter{table}{0}
\renewcommand{\thetable}{S\arabic{table}}

\noindent {\bf Age of the System}\\
{We may estimate the age of the system from the age of PSR~J2229+6114. For a rotation period of $P=51.6\,$ms and spindown rate $\dot{P}=7.8\times 10^{-14}\rm s~s^{-1}$, the characteristic age can be estimated by $\tau_{\rm c}=P/2\dot{P}=10.5\,$kyr. The angular velocity of the pulsar's rotation (denoted by $\Omega$) is generally assumed to evolve as $\dot{\Omega}\propto -\Omega^n$ with $n$ being the so-called braking index. The true age of the pulsar can be given by \citep[e.g.][]{Liu20}
\begin{equation*}
    T_{\rm age}=\left\{
\begin{array}{ll}
2\tau_c{\rm ln}(\frac{P}{P_0}), \quad n=1\\
\frac{2\tau_c}{n-1}\left[1-\left(\frac{P_0}{P}\right)^{n-1} \right], \quad n\neq 1
\end{array}
\right.
\end{equation*}
where $P_0$ is the initial rotation period of the pulsar based on assumption. We can see that the true age of the pulsar is close to its characteristic age only when the pulsar is treated as a magnetic dipole (i.e. $n=3$) with $P_0 \ll P$. The dependence of $T_{\rm age}$ on $n$ and $P_0$ is shown in Fig.~\ref{fig:age}. It may be interesting to indicate that the age of the pulsar could be even shorter than 1\,kyr if the initial rotation period of the pulsar is very close to the present rotation period, e.g., $P_0\geq 49\,$ms. In this case, the SNR would turn out to be a young SNR instead of a middle-age SNR.}\\

\noindent {\bf Break in the Particle Spectrum}\\
{We employ broken power law functions for both electrons and protons. The break in the electron spectrum may be caused by the cooling process. It appears when the cooling timescale is equal to the age of the system (i.e., the dynamical timescale). For the break in the proton spectrum, it may be caused by the energy-dependent diffusive escape of particles and the break energy can also be estimated by equating the escape timescale (i.e., confinement timescale) to the age of the system. The electron cooling timescale is given by $t_{\rm cool}=E_e/\dot{E}_e$ where
\begin{equation*}
    \dot{E}_e=\frac{4}{3}\sigma_Tc\left(\frac{E_e}{m_ec^2}\right)^2\left[\frac{B^2}{8\pi}+\Sigma_i U_i/\left(1+(\frac{kT_iE_e}{m_e^2c^4})^{0.6}\right)^{(1.9/0.6)}\right],
\end{equation*}
where $\sigma_T$ is the Thomson cross section, $m_e$ is the electron mass, $k$ is the Boltzmann constant, $U_i$ and $T_i$ are the energy density and the temperature of radiation field of category $i$, including the CMB, the interstellar IR/Optical/UV radiation fields. The escape timescale is estimated by $t_{\rm esc}\sim R^2/D$ where $R$ is the width of the SNR tail and $D=\eta^{-1}r_gc/3$ is the spatial diffusion coefficient. Relevant timescales are shown in Fig.~\ref{fig:timescale}. The employed break energy for both electron spectrum and the proton spectrum shown in Fig.~\ref{fig:sed} can be interpreted in this way provided an appropriate choice of $\eta$ and $T_{\rm age}$.}\\

\begin{figure}
\centering
\includegraphics[width=0.8\textwidth]{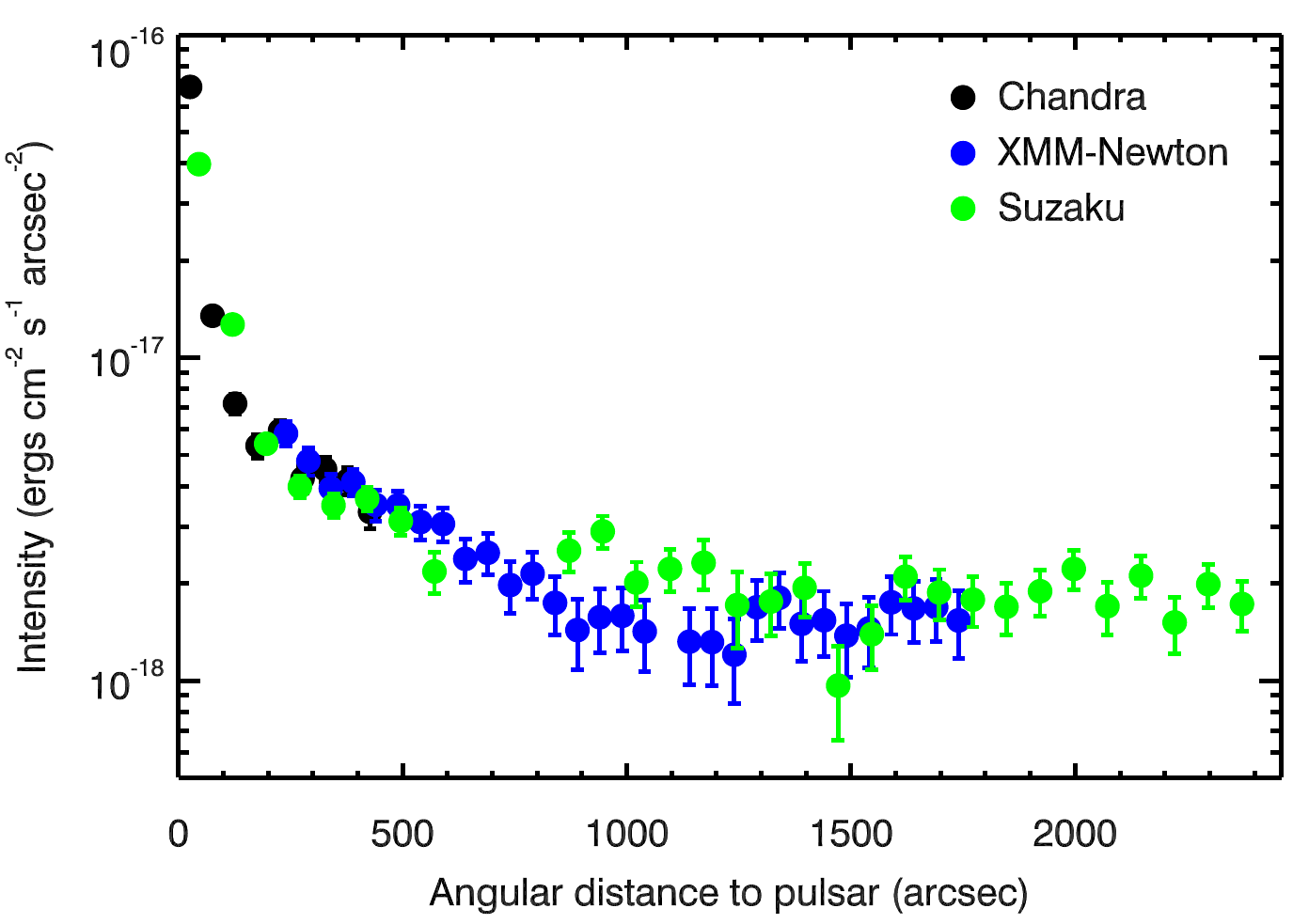}
\caption{{\bf Radial Profiles of Extended X-ray Emission} The profiles from \cha, \xmm, and \suzaku\ are compared. The \cha\ and \xmm\ data are the same with those in Figure~\ref{fig:sbp}A. Note that the \suzaku\ intensity may be more contaminated by unresolved point sources due to its worse angular resolution.}
\label{fig:sbpall}
\end{figure}

\noindent {\bf X-ray Surface Brightness Profile Measured by Suzaku}\\
{The change of SBP in the head and tail regions suggests a different origin of the diffuse X-ray emission in the SNR. The SBP in Fig.~\ref{fig:sbp} is extend to $\sim 30$ arcmin limited by the \xmm\ coverage. The entire SNR is almost covered by the \suzaku\ observations. We also extract a SBP from \suzaku, which extends to a larger radius of $\sim 40$ arcmin in Fig.~\ref{fig:sbpall}. We apply a wider annulus of 75$^{\prime\prime}$ per bin to guarantee a sufficient count statistics of \suzaku\ data in each bin. The obtained \suzaku\ SBP agrees well with the that of \cha\ and \xmm\ in the head region, where the diffuse X-ray emission is dominated by the bright PWN. In the tail region, however, the \suzaku\ SBP are systematically higher than the one from \xmm\ with larger uncertainties due to its limited angular resolution (i.e., it contains the contribution of more unresolved point sources). Nevertheless, the general trend of SBP of \suzaku\ is similar to the one of \cha\ and \xmm: it shows a rapid decline close to the puslar, a shallow decline in the head region and keeps flat in the tail region.}

\clearpage

\begin{figure}
    \centering
    \includegraphics[width=0.8\textwidth]{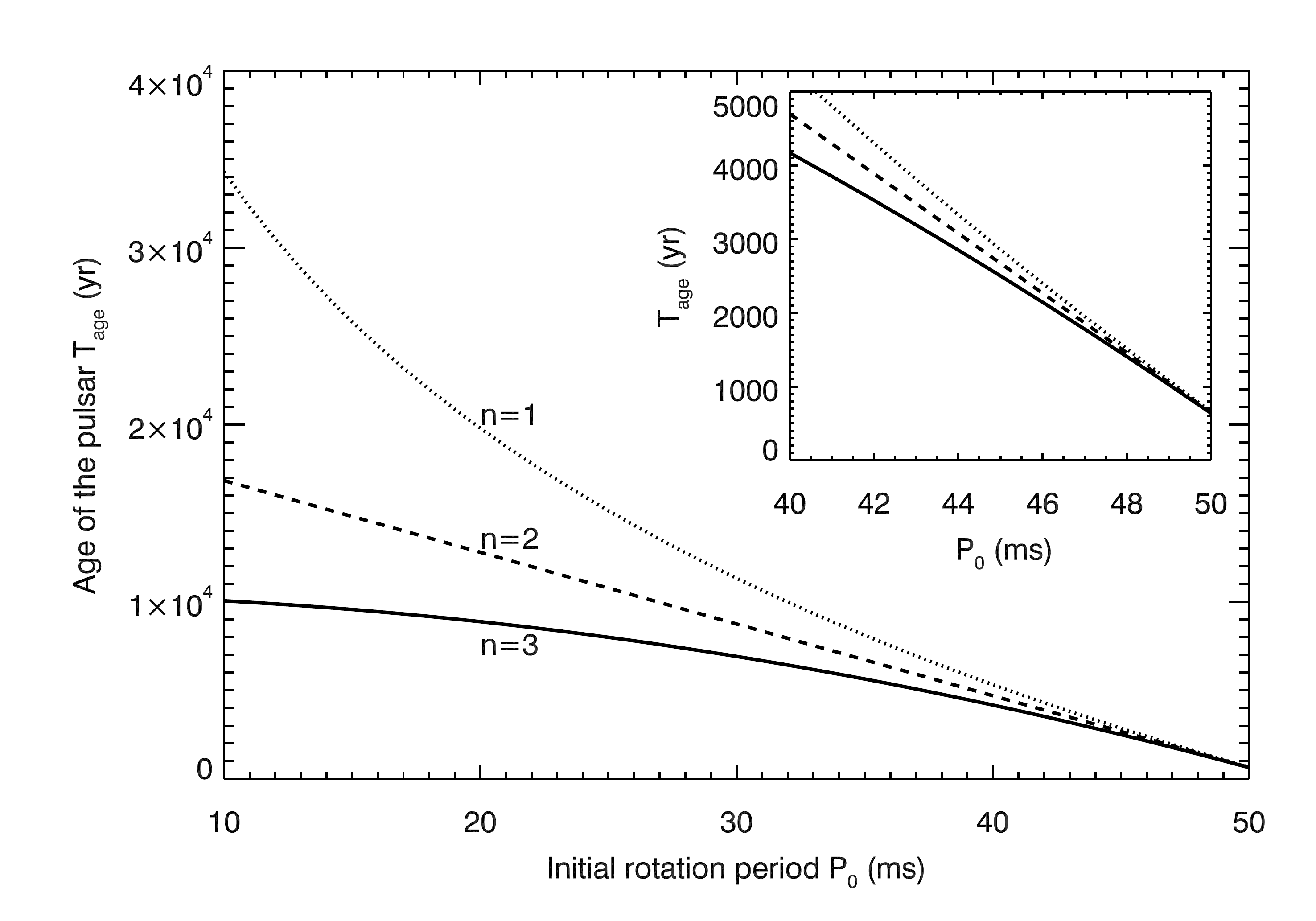}
    \caption{{\bf Age of PSR~J2229+6114} The estimated age of the pulsar is a function of the assumed initial rotation period $P_0$. The solid, dashed, dotted curves represent the result for braking index $n=3,2,1$ respectively. The inset zooms into $P_0=40-50\,$ms.}
    \label{fig:age}
\end{figure}

\clearpage
\begin{figure}
\centering
\includegraphics[width=0.5\textwidth]{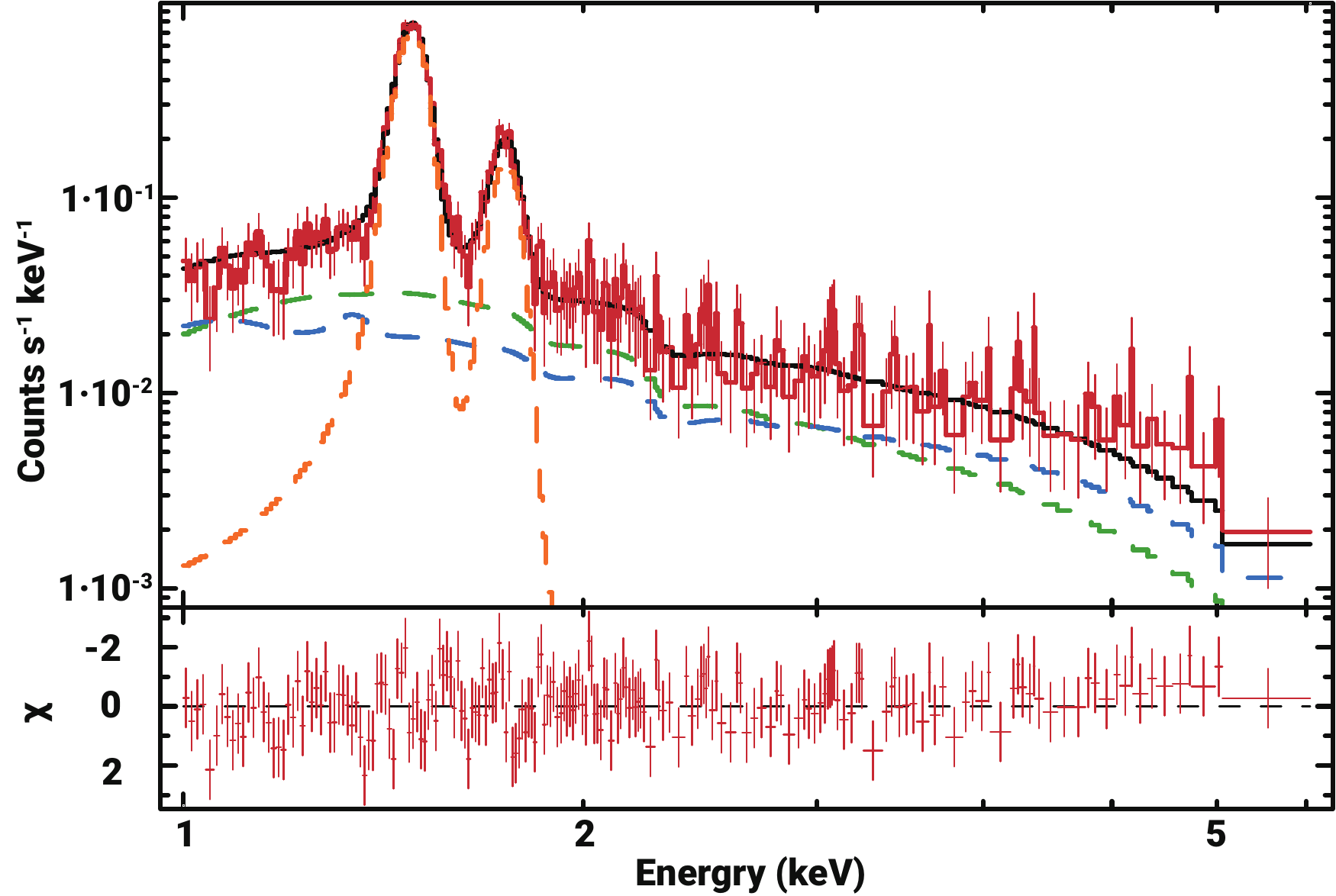}
\includegraphics[width=0.5\textwidth]{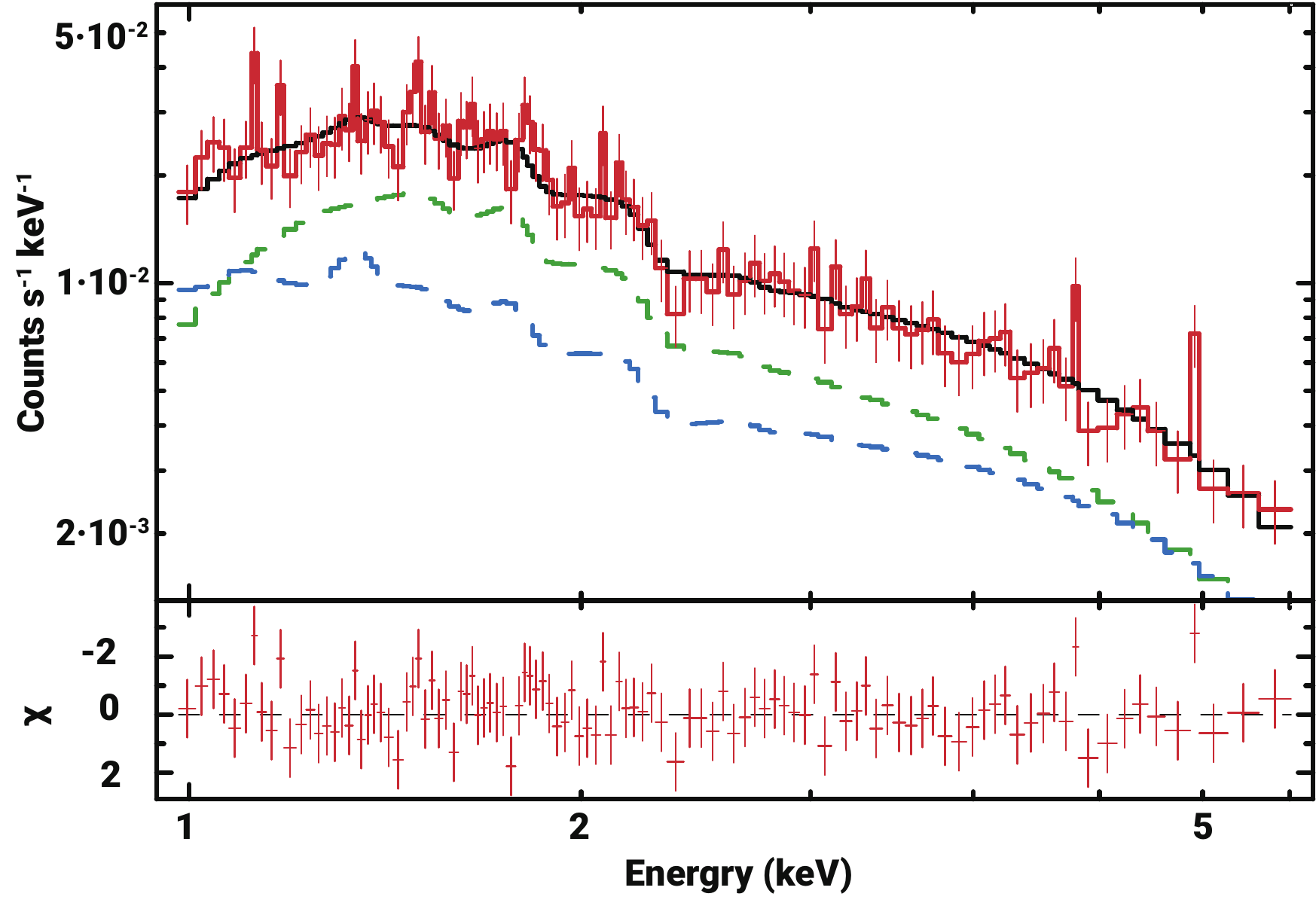}
\includegraphics[width=0.5\textwidth]{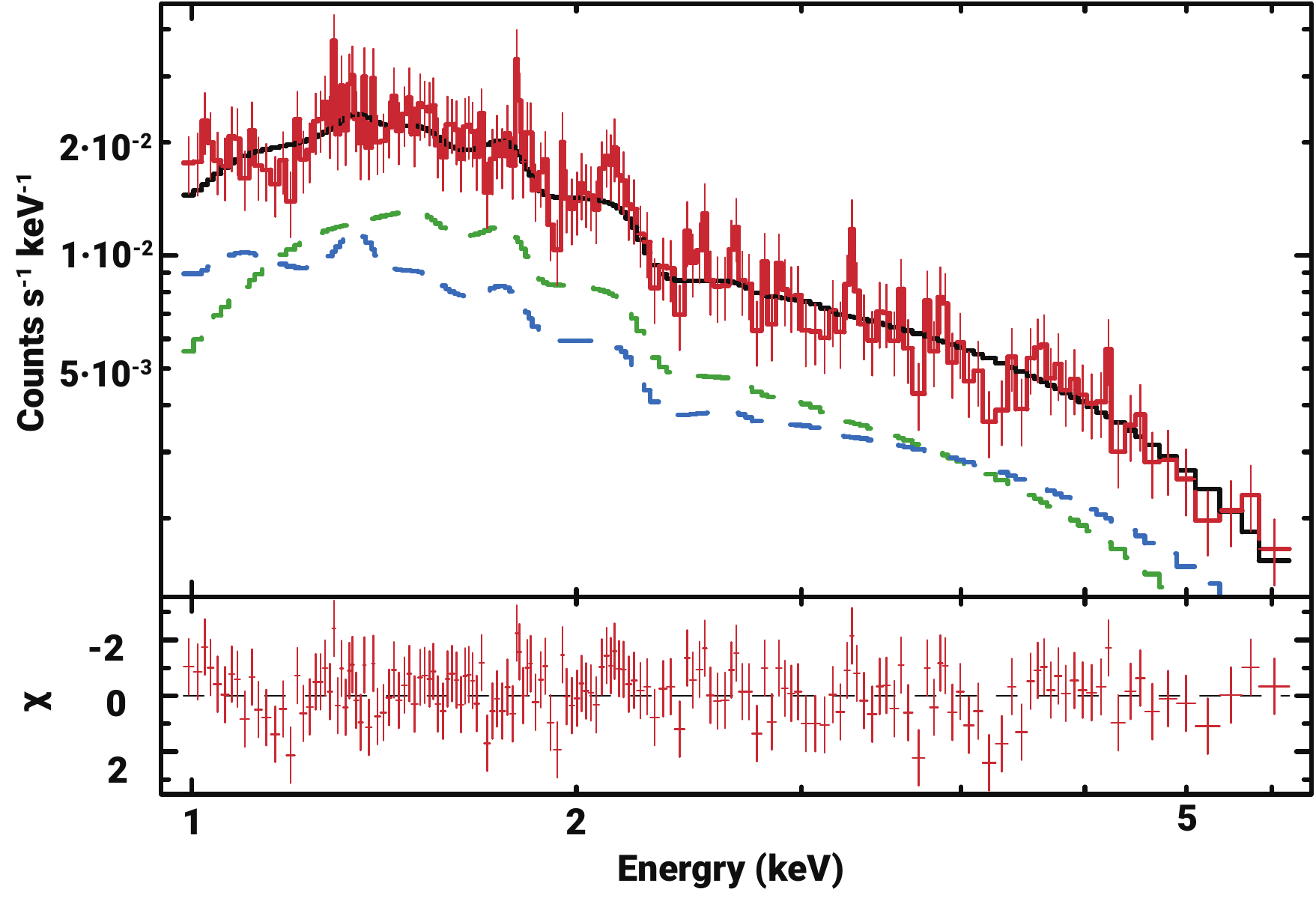}
\caption{{\bf X-ray Spectra} \xmm\ MOS spectra from the TX region (top panel) and \suzaku\ XIS spectra from the TS1/TS2 regions (middle and bottom panels). The red points are data, and the black solid line is the best-fit model. The green dashed lines are a power-law component from the SNR, the blue dashed lines are total components from the sky background. The two orange dashed Gaussian components in \xmm\ spectra are two strong fluorescent instrumental lines of Al K$\alpha$ at 1.49 keV and Si K$\alpha$ at 1.75 keV, which are ignored in the spectral fitting. { These spectra from the tail region do not show any significantly repetitive line features and disfavor the thermal model.}}
\label{fig:spec}
\end{figure}

\begin{figure}
\centering
\includegraphics[width=0.8\textwidth]{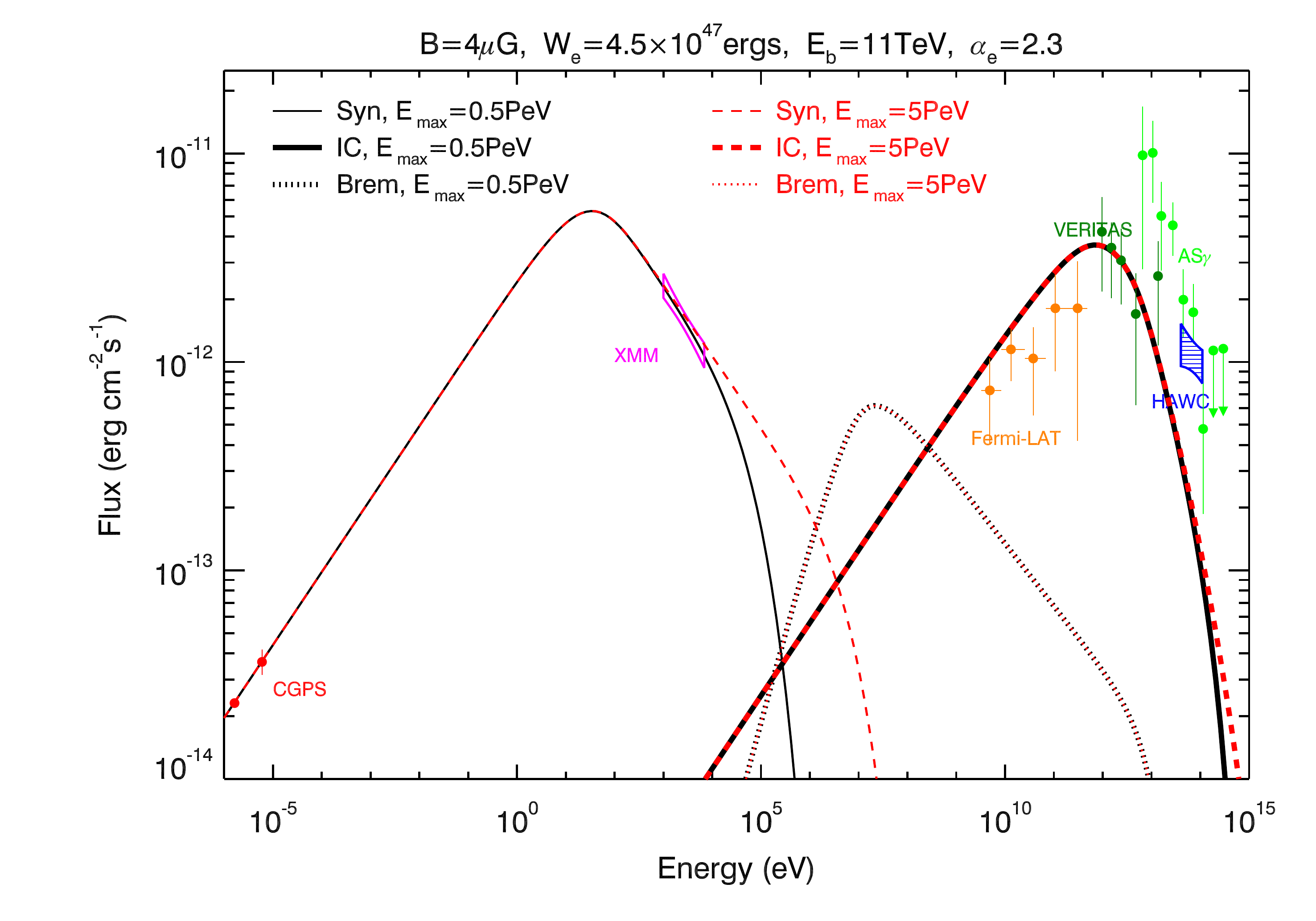}
\caption{{\bf Constraints from the X-ray Data on the IC Interpretation of $>10$\,TeV Flux} The black and red curves represent the predicted flux with $E_{e,\rm max}=0.5\,$PeV and 5\,PeV respectively, while the thin solid, thick solid and dotted curves represent the synchrotron flux, IC flux, and the bremsstrahlung flux respectively. {The gas density for the bremsstrahlung radiation of electrons in the SNR shock is assumed to be $1\,\rm cm^{-3}$}. Other model parameters are labeled on the top of the figure.}
\label{fig:sed_B}
\end{figure}

\begin{figure}
\centering
\includegraphics[width=0.8\textwidth]{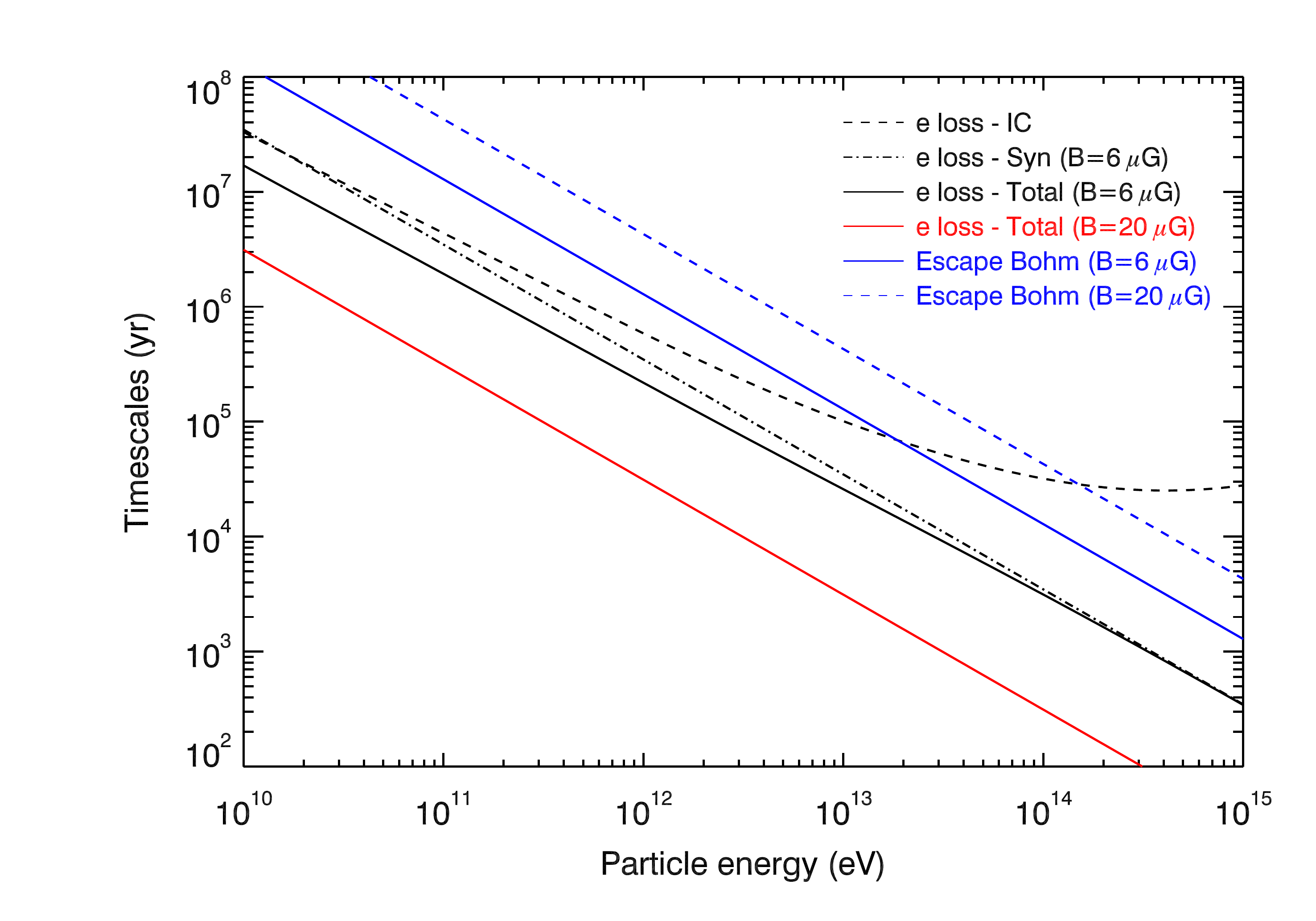}
\caption{{\bf Timescales for Various Processes} The dashed and dash-dotted black curves show the cooling timescale of electrons in a magnetic field of $6\mu$G and in the employed interstellar radiation field (see text for Materials and Method section) respectively, while the solid black curve shows the total cooling timescale. The solid red curve is similar to the solid black one except using a magnetic field of $20\mu$G. The solid and dashed blue curves represent the diffusive escape timescale in the Bohm limit with assuming the size of the source to be 5\,pc.}
\label{fig:timescale}
\end{figure}

\clearpage

\clearpage

\begin{table}[h]
\centering
  \begin{tabular}{@{}lccc@{}}
\hline\hline
Obs-ID & Date & Exp (ks) & Clean Exp (ks) \\  
\hline
\cha\ & & & \\
1948 & 2001-02-14 & 17.7 & 17.7\\
2787 & 2002-03-15 & 94.0 & 93.6\\
\hline
\suzaku\ & & & \\
505054010 & 2010-05-16 & 59.4 & 54.9 \\
505072010 & 2010-08-15 & 24.6 & 24.6 \\
505073010 & 2010-08-16 & 50.5 & 48.6 \\
\hline
\xmm\ & & & \\
0820840301 & 2019-02-06 & 48.6/43.9 & 29.6/16.2\\
0820840401 & 2019-02-10 & 50.9/46.8 & 22.3/15.4\\
0820840501 & 2019-02-14 & 35.3/32.9 & 12.7/5.9\\
\hline
\end{tabular}
  \caption{List of X-ray observations. \suzaku\ exposures are for the clean events of each XIS in 3x3 mode.
\xmm\ exposures are for MOS and pn, respectively.}
\label{t:obs}
\end{table}

\vspace{20pt}

\noindent \textbf{\large Acknowledgements}\\
The authors thank all four anonymous referees for their constructive reports. This work is supported by NSFC Grants No.~U2031105, No.~11625312, No.~11851304 and the
National Key R~\&~D program of China under Grant
No.~2018YFA0404203.\\

\noindent \textbf{\large Author Contributions}\\
R.-Y.L. initiated the project. R.-Y.L and C.G. wrote the manuscript. C.G. analyzed the \cha\ and \xmm\ data. S.N. analyzed the \suzaku\ data. R.-Y.L. led the interpretation of the data. Y.C and X.-Y.W gave advice on data analysis and theoretical interpretation. All the authors discussed the results and commented on the manuscript.\\

\noindent \textbf{\large Declaration of Interests}\\
The authors declare no competing interests.

\bibliographystyle{apj}
\bibliography{ms}

\begin{thebibliography}{50}
\expandafter\ifx\csname natexlab\endcsname\relax\def\natexlab#1{#1}\fi

\bibitem[{{Aartsen} {et~al.}(2017){Aartsen}, {Abraham}, {Ackermann}, {Adams},
  {Aguilar}, {Ahlers}, {Ahrens}, {Altmann}, {Andeen}, {Anderson}, {Ansseau},
  {Anton}, {Archinger}, {Arg{\"u}elles}, {Auffenberg}, {Axani}, {Bai},
  {Barwick}, {Baum}, {Bay}, {Beatty}, {Becker Tjus}, {Becker}, {BenZvi},
  {Berley}, {Bernardini}, {Bernhard}, {Besson}, {Binder}, {Bindig}, {Bissok},
  {Blaufuss}, {Blot}, {Bohm}, {B{\"o}rner}, {Bos}, {Bose}, {B{\"o}ser},
  {Botner}, {Braun}, {Brayeur}, {Bretz}, {Bron}, {Burgman}, {Carver}, {Casier},
  {Cheung}, {Chirkin}, {Christov}, {Clark}, {Classen}, {Coenders}, {Collin},
  {Conrad}, {Cowen}, {Cross}, {Day}, {de Andr{\'e}}, {De Clercq}, {del Pino
  Rosendo}, {Dembinski}, {De Ridder}, {Desiati}, {de Vries}, {de Wasseige}, {de
  With}, {DeYoung}, {D{\'\i}az-V{\'e}lez}, {di Lorenzo}, {Dujmovic}, {Dumm},
  {Dunkman}, {Eberhardt}, {Ehrhardt}, {Eichmann}, {Eller}, {Euler}, {Evenson},
  {Fahey}, {Fazely}, {Feintzeig}, {Felde}, {Filimonov}, {Finley}, {Flis},
  {F{\"o}sig}, {Franckowiak}, {Friedman}, {Fuchs}, {Gaisser}, {Gallagher},
  {Gerhardt}, {Ghorbani}, {Giang}, {Gladstone}, {Glauch}, {Gl{\"u}senkamp},
  {Goldschmidt}, {Golup}, {Gonzalez}, {Grant}, {Griffith}, {Haack}, {Haj
  Ismail}, {Hallgren}, {Halzen}, {Hansen}, {Hansmann}, {Hanson}, {Hebecker},
  {Heereman}, {Helbing}, {Hellauer}, {Hickford}, {Hignight}, {Hill}, {Hoffman},
  {Hoffmann}, {Holzapfel}, {Hoshina}, {Huang}, {Huber}, {Hultqvist}, {In},
  {Ishihara}, {Jacobi}, {Japaridze}, {Jeong}, {Jero}, {Jones}, {Jurkovic},
  {Kappes}, {Karg}, {Karle}, {Katz}, {Kauer}, {Keivani}, {Kelley},
  {Kheirandish}, {Kim}, {Kintscher}, {Kiryluk}, {Kittler}, {Klein}, {Kohnen},
  {Koirala}, {Kolanoski}, {Konietz}, {K{\"o}pke}, {Kopper}, {Kopper},
  {Koskinen}, {Kowalski}, {Krings}, {Kroll}, {Kr{\"u}ckl}, {Kr{\"u}ger},
  {Kunnen}, {Kunwar}, {Kurahashi}, {Kuwabara}, {Labare}, {Lanfranchi},
  {Larson}, {Lauber}, {Lennarz}, {Lesiak-Bzdak}, {Leuermann}, {Lu},
  {L{\"u}nemann}, {Madsen}, {Maggi}, {Mahn}, {Mancina}, {Mandelartz},
  {Maruyama}, {Mase}, {Maunu}, {McNally}, {Meagher}, {Medici}, {Meier}, {Meli},
  {Menne}, {Merino}, {Meures}, {Miarecki}, {Mohrmann}, {Montaruli}, {Moulai},
  {Nahnhauer}, {Naumann}, {Neer}, {Niederhausen}, {Nowicki}, {Nygren},
  {Obertacke Pollmann}, {Olivas}, {O'Murchadha}, {Palczewski}, {Pandya},
  {Pankova}, {Peiffer}, {Penek}, {Pepper}, {P{\'e}rez de los Heros}, {Pieloth},
  {Pinat}, {Price}, {Przybylski}, {Quinnan}, {Raab}, {R{\"a}del}, {Rameez},
  {Rawlins}, {Reimann}, {Relethford}, {Relich}, {Resconi}, {Rhode}, {Richman},
  {Riedel}, {Robertson}, {Rongen}, {Rott}, {Ruhe}, {Ryckbosch}, {Rysewyk},
  {Sabbatini}, {Sanchez Herrera}, {Sandrock}, {Sandroos}, {Sarkar},
  {Satalecka}, {Schlunder}, {Schmidt}, {Schoenen}, {Sch{\"o}neberg},
  {Schumacher}, {Seckel}, {Seunarine}, {Soldin}, {Song}, {Spiczak}, {Spiering},
  {Stanev}, {Stasik}, {Stettner}, {Steuer}, {Stezelberger}, {Stokstad},
  {St{\"o}ssl}, {Str{\"o}m}, {Strotjohann}, {Sullivan}, {Sutherland},
  {Taavola}, {Taboada}, {Tatar}, {Tenholt}, {Ter-Antonyan}, {Terliuk},
  {Te{\v{s}}i{\'c}}, {Tilav}, {Toale}, {Tobin}, {Toscano}, {Tosi},
  {Tselengidou}, {Turcati}, {Unger}, {Usner}, {Vandenbroucke}, {van
  Eijndhoven}, {Vanheule}, {van Rossem}, {van Santen}, {Veenkamp}, {Vehring},
  {Voge}, {Vogel}, {Vraeghe}, {Walck}, {Wallace}, {Wallraff}, {Wandkowsky},
  {Weaver}, {Weiss}, {Wendt}, {Westerhoff}, {Whelan}, {Wickmann}, {Wiebe},
  {Wiebusch}, {Wille}, {Williams}, {Wills}, {Wolf}, {Wood}, {Woolsey},
  {Woschnagg}, {Xu}, {Xu}, {Xu}, {Yanez}, {Yodh}, {Yoshida}, {Zoll}, \&
  {IceCube Collaboration}}]{IC17}
{Aartsen}, M.~G. {et al.}\  2017, \apj, 835, 151

\bibitem[{{Aartsen} {et~al.}(2020){Aartsen}, {Ackermann}, {Adams}, {Aguilar},
  {Ahlers}, {Ahrens}, {Alispach}, {Andeen}, {Anderson}, {Ansseau}, {Anton},
  {Arg{\"u}elles}, {Auffenberg}, {Axani}, {Bagherpour}, {Bai}, {Balagopal V.},
  {Barbano}, {Barwick}, {Bastian}, {Baum}, {Baur}, {Bay}, {Beatty}, {Becker},
  {Becker Tjus}, {BenZvi}, {Berley}, {Bernardini}, {Besson}, {Binder},
  {Bindig}, {Blaufuss}, {Blot}, {Bohm}, {B{\"o}ser}, {Botner}, {B{\"o}ttcher},
  {Bourbeau}, {Bourbeau}, {Bradascio}, {Braun}, {Bron}, {Brostean-Kaiser},
  {Burgman}, {Buscher}, {Busse}, {Carver}, {Chen}, {Cheung}, {Chirkin}, {Choi},
  {Clark}, {Clark}, {Classen}, {Coleman}, {Collin}, {Conrad}, {Coppin},
  {Correa}, {Cowen}, {Cross}, {Dave}, {De Clercq}, {DeLaunay}, {Dembinski},
  {Deoskar}, {De Ridder}, {Desiati}, {de Vries}, {de Wasseige}, {de With},
  {DeYoung}, {Diaz}, {D{\'\i}az-V{\'e}lez}, {Dujmovic}, {Dunkman}, {Dvorak},
  {Eberhardt}, {Ehrhardt}, {Eller}, {Engel}, {Evenson}, {Fahey}, {Fazely},
  {Felde}, {Filimonov}, {Finley}, {Fox}, {Franckowiak}, {Friedman}, {Fritz},
  {Gaisser}, {Gallagher}, {Ganster}, {Garrappa}, {Gerhardt}, {Ghorbani},
  {Glauch}, {Gl{\"u}senkamp}, {Goldschmidt}, {Gonzalez}, {Grant},
  {Gr{\'e}goire}, {Griffith}, {Griswold}, {G{\"u}nder}, {G{\"u}nd{\"u}z},
  {Haack}, {Hallgren}, {Halliday}, {Halve}, {Halzen}, {Hanson}, {Haungs},
  {Hebecker}, {Heereman}, {Heix}, {Helbing}, {Hellauer}, {Henningsen},
  {Hickford}, {Hignight}, {Hill}, {Hoffman}, {Hoffmann}, {Hoinka},
  {Hokanson-Fasig}, {Hoshina}, {Huang}, {Huber}, {Huber}, {Hultqvist},
  {H{\"u}nnefeld}, {Hussain}, {In}, {Iovine}, {Ishihara}, {Jansson},
  {Japaridze}, {Jeong}, {Jero}, {Jones}, {Jonske}, {Joppe}, {Kang}, {Kang},
  {Kappes}, {Kappesser}, {Karg}, {Karl}, {Karle}, {Katz}, {Kauer},
  {Kellermann}, {Kelley}, {Kheirandish}, {Kim}, {Kintscher}, {Kiryluk},
  {Kittler}, {Klein}, {Koirala}, {Kolanoski}, {K{\"o}pke}, {Kopper}, {Kopper},
  {Koskinen}, {Kowalski}, {Krings}, {Kr{\"u}ckl}, {Kulacz}, {Kurahashi},
  {Kyriacou}, {Lanfranchi}, {Larson}, {Lauber}, {Lazar}, {Leonard},
  {Leszczy{\'n}ska}, {Liu}, {Lohfink}, {Lozano Mariscal}, {Lu}, {Lucarelli},
  {Ludwig}, {L{\"u}nemann}, {Luszczak}, {Lyu}, {Ma}, {Madsen}, {Maggi}, {Mahn},
  {Makino}, {Mallik}, {Mallot}, {Mancina}, {Mari{\c{s}}}, {Maruyama}, {Mase},
  {Maunu}, {McNally}, {Meagher}, {Medici}, {Medina}, {Meier}, {Meighen-Berger},
  {Merino}, {Meures}, {Micallef}, {Mockler}, {Moment{\'e}}, {Montaruli},
  {Moore}, {Morse}, {Moulai}, {Muth}, {Nagai}, {Naumann}, {Neer}, {Nguy{\^e}n},
  {Niederhausen}, {Nisa}, {Nowicki}, {Nygren}, {Obertacke Pollmann}, {Oehler},
  {Olivas}, {O'Murchadha}, {O'Sullivan}, {Palczewski}, {Pandya}, {Pankova},
  {Park}, {Peiffer}, {P{\'e}rez de los Heros}, {Philippen}, {Pieloth},
  {Pieper}, {Pinat}, {Pizzuto}, {Plum}, {Porcelli}, {Price}, {Przybylski},
  {Raab}, {Raissi}, {Rameez}, {Rauch}, {Rawlins}, {Rea}, {Rehman}, {Reimann},
  {Relethford}, {Renschler}, {Renzi}, {Resconi}, {Rhode}, {Richman},
  {Robertson}, {Rongen}, {Rott}, {Ruhe}, {Ryckbosch}, {Rysewyk Cantu}, {Safa},
  {Sanchez Herrera}, {Sandrock}, {Sandroos}, {Santander}, {Sarkar}, {Sarkar},
  {Satalecka}, {Schaufel}, {Schieler}, {Schlunder}, {Schmidt}, {Schneider},
  {Schneider}, {Schr{\"o}der}, {Schumacher}, {Sclafani}, {Seckel}, {Seunarine},
  {Shefali}, {Silva}, {Snihur}, {Soedingrekso}, {Soldin}, {Song}, {Spiczak},
  {Spiering}, {Stachurska}, {Stamatikos}, {Stanev}, {Stein}, {Stettner},
  {Steuer}, {Stezelberger}, {Stokstad}, {St{\"o}{\ss}l}, {Strotjohann},
  {St{\"u}rwald}, {Stuttard}, {Sullivan}, {Taboada}, {Tenholt}, {Ter-Antonyan},
  {Terliuk}, {Tilav}, {Tollefson}, {Tomankova}, {T{\"o}nnis}, {Toscano},
  {Tosi}, {Trettin}, {Tselengidou}, {Tung}, {Turcati}, {Turcotte}, {Turley},
  {Ty}, {Unger}, {Unland Elorrieta}, {Usner}, {Vandenbroucke}, {Van Driessche},
  {van Eijk}, {van Eijndhoven}, {van Santen}, {Verpoest}, {Vraeghe}, {Walck},
  {Wallace}, {Wallraff}, {Wandkowsky}, {Watson}, {Weaver}, {Weindl}, {Weiss},
  {Weldert}, {Wendt}, {Werthebach}, {Whelan}, {Whitehorn}, {Wiebe}, {Wiebusch},
  {Wille}, {Williams}, {Wills}, {Wolf}, {Wood}, {Wood}, {Woschnagg}, {Wrede},
  {Xu}, {Xu}, {Xu}, {Yanez}, {Yodh}, {Yoshida}, {Yuan}, {Z{\"o}cklein}, \&
  {IceCube Collaboration}}]{IC20_PWN}
--- 2020, \apj, 898, 117

\bibitem[{{Abdo} {et~al.}(2007){Abdo}, {Allen}, {Berley}, {Casanova}, {Chen},
  {Coyne}, {Dingus}, {Ellsworth}, {Fleysher}, {Fleysher}, {Gonzalez},
  {Goodman}, {Hays}, {Hoffman}, {Hopper}, {H{\"u}ntemeyer}, {Kolterman},
  {Lansdell}, {Linnemann}, {McEnery}, {Mincer}, {Nemethy}, {Noyes}, {Ryan},
  {Saz Parkinson}, {Shoup}, {Sinnis}, {Smith}, {Sullivan}, {Vasileiou},
  {Walker}, {Williams}, {Xu}, \& {Yodh}}]{Milagro07}
{Abdo}, A.~A. {et al.}\  2007, \apjl, 664, L91

\bibitem[{{Abdo} {et~al.}(2009){Abdo}, {Allen}, {Aune}, {Berley}, {Chen},
  {Christopher}, {DeYoung}, {Dingus}, {Ellsworth}, {Gonzalez}, {Goodman},
  {Hays}, {Hoffman}, {H{\"u}ntemeyer}, {Kolterman}, {Linnemann}, {McEnery},
  {Morgan}, {Mincer}, {Nemethy}, {Pretz}, {Ryan}, {Saz Parkinson}, {Shoup},
  {Sinnis}, {Smith}, {Vasileiou}, {Walker}, {Williams}, \& {Yodh}}]{Milagro09}
--- 2009, \apjl, 700, L127

\bibitem[{{Acciari} {et~al.}(2009){Acciari}, {Aliu}, {Arlen}, {Aune},
  {Bautista}, {Beilicke}, {Benbow}, {Boltuch}, {Bradbury}, {Buckley}, {Bugaev},
  {Butt}, {Byrum}, {Cannon}, {Cesarini}, {Chow}, {Ciupik}, {Cogan}, {Cui},
  {Dickherber}, {Ergin}, {Fegan}, {Finley}, {Fortin}, {Fortson}, {Furniss},
  {Gall}, {Gillanders}, {Gotthelf}, {Grube}, {Guenette}, {Gyuk}, {Hanna},
  {Holder}, {Horan}, {Hui}, {Humensky}, {Kaaret}, {Karlsson}, {Kertzman},
  {Kieda}, {Konopelko}, {Krawczynski}, {Krennrich}, {Lang}, {LeBohec}, {Maier},
  {McCann}, {McCutcheon}, {Millis}, {Moriarty}, {Mukherjee}, {Ong}, {Otte},
  {Pandel}, {Perkins}, {Pohl}, {Quinn}, {Ragan}, {Reyes}, {Reynolds}, {Roache},
  {Rose}, {Schroedter}, {Sembroski}, {Smith}, {Steele}, {Swordy}, {Theiling},
  {Toner}, {Vassiliev}, {Vincent}, {Wagner}, {Wakely}, {Ward}, {Weekes},
  {Weinstein}, {Weisgarber}, {Williams}, {Wissel}, {Wood}, \&
  {Zitzer}}]{Veritas09}
{Acciari}, V.~A. {et al.}\  2009, \apjl, 703, L6

\bibitem[{{Akaike}(1974)}]{AIC74}
{Akaike}, H. 1974, IEEE Transactions on Automatic Control, 19, 716

\bibitem[{{Albert} {et~al.}(2020){Albert}, {Alfaro}, {Alvarez}, {Camacho},
  {Arteaga-Vel{\'a}zquez}, {Arunbabu}, {Avila Rojas}, {Ayala Solares},
  {Baghmanyan}, {Belmont-Moreno}, {BenZvi}, {Brisbois}, {Caballero-Mora},
  {Capistr{\'a}n}, {Carrami{\~n}ana}, {Casanova}, {Cotti}, {Cotzomi},
  {Couti{\~n}o de Le{\'o}n}, {De la Fuente}, {Diaz-Cruz}, {Dingus},
  {DuVernois}, {D{\'\i}az-V{\'e}lez}, {Ellsworth}, {Engel}, {Espinoza}, {Fan},
  {Fang}, {Fern{\'a}ndez Alonso}, {Fleischhack}, {Fraija},
  {Galv{\'a}n-G{\'a}mez}, {Garcia}, {Garc{\'\i}a-Gonz{\'a}lez}, {Garfias},
  {Giacinti}, {Gonz{\'a}lez}, {Goodman}, {Harding}, {Hernandez}, {Hinton},
  {Hona}, {Huang}, {Hueyotl-Zahuantitla}, {H{\"u}ntemeyer}, {Iriarte},
  {Jardin-Blicq}, {Joshi}, {Lee}, {Le{\'o}n Vargas}, {Linnemann}, {Longinotti},
  {Luis-Raya}, {Lundeen}, {Malone}, {Marinelli}, {Martinez},
  {Martinez-Castellanos}, {Mart{\'\i}nez-Castro}, {Matthews}, {Mirand
  a-Romagnoli}, {Morales-Soto}, {Moreno}, {Mostaf{\'a}}, {Nayerhoda}, {Nellen},
  {Newbold}, {Nisa}, {Noriega-Papaqui}, {Omodei}, {Peisker}, {P{\'e}rez
  Araujo}, {P{\'e}rez-P{\'e}rez}, {Rho}, {Rosa-Gonz{\'a}lez}, {Ruiz-Velasco},
  {Salazar}, {Salesa Greus}, {Sandoval}, {Schneider}, {Schoorlemmer}, {Serna
  Franco}, {Sinnis}, {Smith}, {Springer}, {Surajbali}, {Tabachnick}, {Tanner},
  {Tibolla}, {Tollefson}, {Torres}, {Torres-Escobedo}, {Ure{\~n}a-Mena},
  {Villase{\~n}or}, {Weisgarber}, {Zepeda}, {Zhou}, {de Le{\'o}n},
  {{\'A}lvarez}, \& {HAWC Collaboration}}]{HAWC20}
{Albert}, A. {et al.}\  2020, \apjl, 896, L29

\bibitem[{{Bai} {et~al.}(2019){Bai}, {Bi}, {Bi}, {Cao}, {Chen}, {Chen},
  {Chiavassa}, {Cui}, {Dai}, {della Volpe}, {Di Girolamo}, {Di Sciascio},
  {Fan}, {Giacalone}, {Guo}, {He}, {He}, {Heller}, {Huang}, {Huang}, {Jia},
  {Ksenofontov}, {Leahy}, {Li}, {Li}, {Liang}, {Lipari}, {Liu}, {Liu}, {Liu},
  {Ma}, {Martineau-Huynh}, {Martraire}, {Montaruli}, {Ruffolo}, {Stenkin},
  {Su}, {Tam}, {Tang}, {Tian}, {Vallania}, {Vernetto}, {Vigorito}, {Wang},
  {Wang}, {Wang}, {Wang}, {Wang}, {Wang}, {Wei}, {Wei}, {Wu}, {Wu}, {Wu},
  {Yan}, {Yang}, {Yang}, {Yao}, {Yin}, {Yuan}, {Zhang}, {Zhang}, {Zhang},
  {Zhang}, {Zhang}, {Zhang}, {Zhao}, {Zhou}, {Zhu}, \& {Zhu}}]{LHAASO19}
{Bai}, X. {et al.}\  2019, arXiv e-prints, arXiv:1905.02773

\bibitem[{{Bamba} {et~al.}(2000){Bamba}, {Koyama}, \& {Tomida}}]{Bamba00}
{Bamba}, A., {Koyama}, K., \& {Tomida}, H. 2000, \pasj, 52, 1157

\bibitem[{{Beck}(2015)}]{Beck15}
{Beck}, R. 2015, \aapr, 24, 4

\bibitem[{{Bell}(1978)}]{Bell78}
{Bell}, A.~R. 1978, \mnras, 182, 147

\bibitem[{{Bell}(2015)}]{Bell15}
--- 2015, \mnras, 447, 2224

\bibitem[{{Bell} {et~al.}(2013){Bell}, {Schure}, {Reville}, \&
  {Giacinti}}]{Bell13}
{Bell}, A.~R., {Schure}, K.~M., {Reville}, B., \& {Giacinti}, G. 2013, \mnras,
  431, 415

\bibitem[{{Blandford} \& {Eichler}(1987)}]{Blandford87}
{Blandford}, R. \& {Eichler}, D. 1987, \physrep, 154, 1

\bibitem[{{Chen} {et~al.}(2006){Chen}, {Wang}, {Gotthelf}, {Jiang}, {Chu}, \&
  {Gruendl}}]{Chen06}
{Chen}, Y., {Wang}, Q.~D., {Gotthelf}, E.~V., {Jiang}, B., {Chu}, Y.-H., \&
  {Gruendl}, R. 2006, \apj, 651, 237

\bibitem[{{Ge} {et~al.}(2020){Ge}, {Liu}, {Sun}, {Yu}, {Rudnick}, {Eilek},
  {Owen}, {Dasadia}, {Rossetti}, {Markevitch}, {Clarke}, {Jones}, {Ghizzardi},
  {Venturi}, {Finoguenov}, \& {Eckert}}]{2020MNRAS.497.4704G}
{Ge}, C. {et al.}\  2020, \mnras, 497, 4704

\bibitem[{{Ge} {et~al.}(2019){Ge}, {Sun}, {Rozo}, {Sehgal}, {Vikhlinin},
  {Forman}, {Jones}, \& {Nagai}}]{2019MNRAS.484.1946G}
{Ge}, C., {Sun}, M., {Rozo}, E., {Sehgal}, N., {Vikhlinin}, A., {Forman}, W.,
  {Jones}, C., \& {Nagai}, D. 2019, \mnras, 484, 1946

\bibitem[{{Gotthelf} {et~al.}(2001){Gotthelf}, {Koralesky}, {Rudnick}, {Jones},
  {Hwang}, \& {Petre}}]{Gotthelf01}
{Gotthelf}, E.~V., {Koralesky}, B., {Rudnick}, L., {Jones}, T.~W., {Hwang}, U.,
  \& {Petre}, R. 2001, \apjl, 552, L39

\bibitem[{{Halpern} {et~al.}(2001{\natexlab{a}}){Halpern}, {Camilo},
  {Gotthelf}, {Helfand }, {Kramer}, {Lyne}, {Leighly}, \&
  {Eracleous}}]{Halpern01b}
{Halpern}, J.~P., {Camilo}, F., {Gotthelf}, E.~V., {Helfand }, D.~J., {Kramer},
  M., {Lyne}, A.~G., {Leighly}, K.~M., \& {Eracleous}, M. 2001{\natexlab{a}},
  \apjl, 552, L125

\bibitem[{{Halpern} {et~al.}(2001{\natexlab{b}}){Halpern}, {Gotthelf},
  {Leighly}, \& {Helfand }}]{Halpern01a}
{Halpern}, J.~P., {Gotthelf}, E.~V., {Leighly}, K.~M., \& {Helfand }, D.~J.
  2001{\natexlab{b}}, \apj, 547, 323

\bibitem[{{Han}(2017)}]{Han17}
{Han}, J.~L. 2017, \araa, 55, 111

\bibitem[{{HESS Collaboration} {et~al.}(2016){HESS Collaboration},
  {Abramowski}, {Aharonian}, {Benkhali}, {Akhperjanian}, {Ang{\"u}ner},
  {Backes}, {Balzer}, {Becherini}, {Tjus}, {Berge}, {Bernhard}, {Bernl{\"o}hr},
  {Birsin}, {Blackwell}, {B{\"o}ttcher}, {Boisson}, {Bolmont}, {Bordas},
  {Bregeon}, {Brun}, {Brun}, {Bryan}, {Bulik}, {Carr}, {Casanova},
  {Chakraborty}, {Chalme-Calvet}, {Chaves}, {Chen}, {Chr{\'e}tien},
  {Colafrancesco}, {Cologna}, {Conrad}, {Couturier}, {Cui}, {Davids},
  {Degrange}, {Deil}, {Dewilt}, {Djannati-Ata{\"\i}}, {Domainko}, {Donath},
  {Drury}, {Dubus}, {Dutson}, {Dyks}, {Dyrda}, {Edwards}, {Egberts}, {Eger},
  {Ernenwein}, {Espigat}, {Farnier}, {Fegan}, {Feinstein}, {Fernandes},
  {Fernand ez}, {Fiasson}, {Fontaine}, {F{\"o}rster}, {F{\"u}{\ss}ling},
  {Gabici}, {Gajdus}, {Gallant}, {Garrigoux}, {Giavitto}, {Giebels},
  {Glicenstein}, {Gottschall}, {Goyal}, {Grondin}, {Grudzi{\'n}ska}, {Hadasch},
  {H{\"a}ffner}, {Hahn}, {Hawkes}, {Heinzelmann}, {Henri}, {Hermann}, {Hervet},
  {Hillert}, {Hinton}, {Hofmann}, {Hofverberg}, {Hoischen}, {Holler}, {Horns},
  {Ivascenko}, {Jacholkowska}, {Jamrozy}, {Janiak}, {Jankowsky},
  {Jung-Richardt}, {Kastendieck}, {Katarzy{\'n}ski}, {Katz}, {Kerszberg},
  {Kh{\'e}lifi}, {Kieffer}, {Klepser}, {Klochkov}, {Klu{\'z}niak}, {Kolitzus},
  {Komin}, {Kosack}, {Krakau}, {Krayzel}, {Kr{\"u}ger}, {Laffon}, {Lamanna},
  {Lau}, {Lefaucheur}, {Lefranc}, {Lemi{\'e}re}, {Lemoine-Goumard}, {Lenain},
  {Lohse}, {Lopatin}, {Lu}, {Lui}, {Marand on}, {Marcowith}, {Mariaud}, {Marx},
  {Maurin}, {Maxted}, {Mayer}, {Meintjes}, {Menzler}, {Meyer}, {Mitchell},
  {Moderski}, {Mohamed}, {Mor{\r{a}}}, {Moulin}, {Murach}, {de Naurois},
  {Niemiec}, {Oakes}, {Odaka}, {{\"O}ttl}, {Ohm}, {Opitz}, {Ostrowski}, {Oya},
  {Panter}, {Parsons}, {Arribas}, {Pekeur}, {Pelletier}, {Petrucci}, {Peyaud},
  {Pita}, {Poon}, {Prokoph}, {P{\"u}hlhofer}, {Punch}, {Quirrenbach}, {Raab},
  {Reichardt}, {Reimer}, {Reimer}, {Renaud}, {de Los Reyes}, {Rieger},
  {Romoli}, {Rosier-Lees}, {Rowell}, {Rudak}, {Rulten}, {Sahakian}, {Salek},
  {Sanchez}, {Santangelo}, {Sasaki}, {Schlickeiser}, {Sch{\"u}ssler}, {Schulz},
  {Schwanke}, {Schwemmer}, {Seyffert}, {Simoni}, {Sol}, {Spanier}, {Spengler},
  {Spies}, {Stawarz}, {Steenkamp}, {Stegmann}, {Stinzing}, {Stycz}, {Sushch},
  {Tavernet}, {Tavernier}, {Taylor}, {Terrier}, {Tluczykont}, {Trichard},
  {Tuffs}, {Valerius}, {van der Walt}, {van Eldik}, {van Soelen},
  {Vasileiadis}, {Veh}, {Venter}, {Viana}, {Vincent}, {Vink}, {Voisin},
  {V{\"o}lk}, {Vuillaume}, {Wagner}, {Wagner}, {Wagner}, {Weidinger},
  {Weitzel}, {White}, {Wierzcholska}, {Willmann}, {W{\"o}rnlein}, {Wouters},
  {Yang}, {Zabalza}, {Zaborov}, {Zacharias}, {Zdziarski}, {Zech}, {Zefi}, \&
  {{\.Z}ywucka}}]{HESS16_GC}
{HESS Collaboration} {et al.}\  2016, \nat, 531, 476

\bibitem[{{Hwang} {et~al.}(2002){Hwang}, {Decourchelle}, {Holt}, \&
  {Petre}}]{Hwang02}
{Hwang}, U., {Decourchelle}, A., {Holt}, S.~S., \& {Petre}, R. 2002, \apj, 581,
  1101

\bibitem[{{Joncas} \& {Higgs}(1990)}]{Joncas90}
{Joncas}, G. \& {Higgs}, L.~A. 1990, \aaps, 82, 113

\bibitem[{{Kafexhiu} {et~al.}(2014){Kafexhiu}, {Aharonian}, {Taylor}, \&
  {Vila}}]{Kafexhiu14}
{Kafexhiu}, E., {Aharonian}, F., {Taylor}, A.~M., \& {Vila}, G.~S. 2014, \prd,
  90, 123014

\bibitem[{{Kirk} \& {Dendy}(2001)}]{Kirk01}
{Kirk}, J.~G. \& {Dendy}, R.~O. 2001, Journal of Physics G Nuclear Physics, 27,
  1589

\bibitem[{{Kothes} {et~al.}(2006){Kothes}, {Reich}, \&
  {Uyan{\i}ker}}]{Kothes06}
{Kothes}, R., {Reich}, W., \& {Uyan{\i}ker}, B. 2006, \apj, 638, 225

\bibitem[{{Kothes} {et~al.}(2001){Kothes}, {Uyaniker}, \&
  {Pineault}}]{Kothes01}
{Kothes}, R., {Uyaniker}, B., \& {Pineault}, S. 2001, \apj, 560, 236

\bibitem[{{Koyama} {et~al.}(1997){Koyama}, {Kinugasa}, {Matsuzaki},
  {Nishiuchi}, {Sugizaki}, {Torii}, {Yamauchi}, \& {Aschenbach}}]{Koyama97}
{Koyama}, K., {Kinugasa}, K., {Matsuzaki}, K., {Nishiuchi}, M., {Sugizaki}, M.,
  {Torii}, K., {Yamauchi}, S., \& {Aschenbach}, B. 1997, \pasj, 49, L7

\bibitem[{{Koyama} {et~al.}(1995){Koyama}, {Petre}, {Gotthelf}, {Hwang},
  {Matsuura}, {Ozaki}, \& {Holt}}]{Koyama95}
{Koyama}, K., {Petre}, R., {Gotthelf}, E.~V., {Hwang}, U., {Matsuura}, M.,
  {Ozaki}, M., \& {Holt}, S.~S. 1995, \nat, 378, 255

\bibitem[{{Liu} \& {Yan}(2020)}]{Liu20}
{Liu}, R.-Y. \& {Yan}, H. 2020, \mnras, 494, 2618

\bibitem[{{Liu} {et~al.}(2020){Liu}, {Zeng}, {Xin}, \& {Zhu}}]{LiuSM20}
{Liu}, S., {Zeng}, H., {Xin}, Y., \& {Zhu}, H. 2020, \apjl, 897, L34

\bibitem[{{Ohira} \& {Yamazaki}(2017)}]{Ohira17}
{Ohira}, Y. \& {Yamazaki}, R. 2017, Journal of High Energy Astrophysics, 13, 17

\bibitem[{{Parizot} {et~al.}(2006){Parizot}, {Marcowith}, {Ballet}, \&
  {Gallant}}]{Parizot06}
{Parizot}, E., {Marcowith}, A., {Ballet}, J., \& {Gallant}, Y.~A. 2006, \aap,
  453, 387

\bibitem[{{Pineault} \& {Joncas}(2000)}]{Pineault00}
{Pineault}, S. \& {Joncas}, G. 2000, \aj, 120, 3218

\bibitem[{{Popescu} {et~al.}(2017){Popescu}, {Yang}, {Tuffs}, {Natale},
  {Rushton}, \& {Aharonian}}]{Popescu17}
{Popescu}, C.~C., {Yang}, R., {Tuffs}, R.~J., {Natale}, G., {Rushton}, M., \&
  {Aharonian}, F. 2017, \mnras, 470, 2539

\bibitem[{{Reynolds} {et~al.}(2008){Reynolds}, {Borkowski}, {Green}, {Hwang},
  {Harrus}, \& {Petre}}]{Reynolds08}
{Reynolds}, S.~P., {Borkowski}, K.~J., {Green}, D.~A., {Hwang}, U., {Harrus},
  I., \& {Petre}, R. 2008, \apjl, 680, L41

\bibitem[{{Reynolds} {et~al.}(2007){Reynolds}, {Borkowski}, {Hwang}, {Hughes},
  {Badenes}, {Laming}, \& {Blondin}}]{Reynolds07}
{Reynolds}, S.~P., {Borkowski}, K.~J., {Hwang}, U., {Hughes}, J.~P., {Badenes},
  C., {Laming}, J.~M., \& {Blondin}, J.~M. 2007, \apjl, 668, L135

\bibitem[{{Rieger} {et~al.}(2007){Rieger}, {Bosch-Ramon}, \&
  {Duffy}}]{Rieger07}
{Rieger}, F.~M., {Bosch-Ramon}, V., \& {Duffy}, P. 2007, \apss, 309, 119

\bibitem[{{Schure} \& {Bell}(2013)}]{Schure13}
{Schure}, K.~M. \& {Bell}, A.~R. 2013, \mnras, 435, 1174

\bibitem[{{Slane} {et~al.}(1999){Slane}, {Gaensler}, {Dame}, {Hughes},
  {Plucinsky}, \& {Green}}]{Slane97}
{Slane}, P., {Gaensler}, B.~M., {Dame}, T.~M., {Hughes}, J.~P., {Plucinsky},
  P.~P., \& {Green}, A. 1999, \apj, 525, 357

\bibitem[{{Slane} {et~al.}(2004){Slane}, {Helfand}, {van der Swaluw}, \&
  {Murray}}]{Slane04}
{Slane}, P., {Helfand}, D.~J., {van der Swaluw}, E., \& {Murray}, S.~S. 2004,
  \apj, 616, 403

\bibitem[{{Tibet AS{\ensuremath{\gamma}} Collaboration} {et~al.}(2021){Tibet
  AS{\ensuremath{\gamma}} Collaboration}, {Amenomori}, {Bao}, {Bi}, {Chen},
  {Chen}, {Chen}, {Chen}, {Chen}, {Cirennima}, {Danzengluobu}, {Fang}, {Fang},
  {Feng}, {Feng}, {Feng}, {Gao}, {Gou}, {Guo}, {Guo}, {He}, {He}, {Hibino},
  {Hotta}, {Hu}, {Hu}, {Huang}, {Jia}, {Jiang}, {Jin}, {Kasahara}, {Katayose},
  {Kato}, {Kato}, {Kawata}, {Kihara}, {Ko}, {Kozai}, {Labaciren}, {Li}, {Li},
  {Li}, {Lin}, {Liu}, {Liu}, {Liu}, {Liu}, {Liu}, {Lou}, {Lu}, {Meng},
  {Munakata}, {Nakada}, {Nakamura}, {Nanjo}, {Nishizawa}, {Ohnishi}, {Ohura},
  {Ozawa}, {Qian}, {Qu}, {Saito}, {Sakata}, {Sako}, {Shao}, {Shibata},
  {Shiomi}, {Sugimoto}, {Takano}, {Takita}, {Tan}, {Tateyama}, {Torii},
  {Tsuchiya}, {Udo}, {Wang}, {Wu}, {Xue}, {Yamamoto}, {Yang}, {Yokoe}, {Yuan},
  {Zhai}, {Zhang}, {Zhang}, {Zhang}, {Zhang}, {Zhang}, {Zhang}, {Zhang},
  {Zhao}, \& {Zhaxisangzhu}}]{2021NatAs.tmp...41T}
{Tibet AS{\ensuremath{\gamma}} Collaboration} {et al.}\  2021, Nature Astronomy

\bibitem[{{Tsujimoto} {et~al.}(2011){Tsujimoto}, {Guainazzi}, {Plucinsky},
  {Beardmore}, {Ishida}, {Natalucci}, {Posson-Brown}, {Read}, {Saxton}, \&
  {Shaposhnikov}}]{Tsujimoto11}
{Tsujimoto}, M. {et al.}\  2011, \aap, 525, A25

\bibitem[{{Van Etten} \& {Romani}(2011)}]{vanEtten11}
{Van Etten}, A. \& {Romani}, R.~W. 2011, \apj, 742, 62

\bibitem[{{Voelk} \& {Biermann}(1988)}]{Voelk88}
{Voelk}, H.~J. \& {Biermann}, P.~L. 1988, \apjl, 333, L65

\bibitem[{{Xin} {et~al.}(2019){Xin}, {Zeng}, {Liu}, {Fan}, \& {Wei}}]{Xin19}
{Xin}, Y., {Zeng}, H., {Liu}, S., {Fan}, Y., \& {Wei}, D. 2019, \apj, 885, 162

\bibitem[{{Zhang} \& {Liu}(2019)}]{ZhangX19}
{Zhang}, X. \& {Liu}, S. 2019, \apj, 876, 24

\bibitem[{{Zirakashvili} \& {Aharonian}(2007)}]{Zirakashvili07}
{Zirakashvili}, V.~N. \& {Aharonian}, F. 2007, \aap, 465, 695

\bibitem[{{Zirakashvili} \& {Ptuskin}(2018)}]{Zirakashvili18}
{Zirakashvili}, V.~N. \& {Ptuskin}, V.~S. 2018, Astroparticle Physics, 98, 21

\end{thebibliography}

\end{document}